\documentclass[epj]{svjour}
\usepackage[dvips]{graphicx}
\begin{document}
\title{Power laws in elementary and heavy-ion collisions}
\subtitle{- A story of fluctuations and nonextensivity? -}
\author{Grzegorz Wilk \inst{1}
\thanks{\emph{e-mail: wilk@fuw.edu.pl}}
\and Zbigniew W\l odarczyk\inst{2}
\thanks{\emph{e-mail: wlod@pu.kielce.pl}}
}                     
\institute{The Andrzej So\l tan Institute for Nuclear Studies,
                Ho\.za 69; 00-689 Warsaw, Poland \and Institute of Physics,
                Jan Kochanowski University,  \'Swi\c{e}tokrzyska 15,
                25-406 Kielce, Poland}
\date{Received: date / Revised version: date}
%
\abstract{ We review from the point of view of nonextensive
statistics the ubiquitous presence in elementary and heavy-ion
collisions of power-law distributions. Special emphasis is placed
on the conjecture that this is just a reflection of some intrinsic
fluctuations existing in the hadronic systems considered. These
systems summarily described by a single parameter $q$ playing the
role of a nonextensivity measure in the nonextensive statistical
models based on Tsallis entropy.
\PACS{
      {89.75.-k}{complex systems}   \and
      {24.60.-k}{statistical theory and fluctuations} \and
      {25.75.Dw}{particle production (relativistic collisions)} \and
      {25.75.-q}{relativistic heavy ions collision}
     }
} 
\maketitle
\section{Introduction}
\label{sec:intro}

In many domains of physics, especially in elementary and heavy-ion
collisions, for decades the prevailing understanding was that
exponential shape of most of the observed spectra of produced
secondaries suggests their statistical (or even thermodynamical)
origin. Therefore, when looking for spectra of transverse momenta
it is commonly assumed that the observed inverse mean transverse
momenta (characterizing the widths of such exponential spectra)
play the role of a temperature of the hadronizing system
\cite{Thermodynamics}. This assumption allows us to use, in what
follows, the whole machinery of statistical models. This is
specially important in the case of high energy collisions of heavy
ions because it allows us to use the tools of statistical physics
to investigate the possibility of the formation in such collisions
of a new state of matter, the so called {\it Quark Gluon Plasma}
(QGP) \cite{QGP} (in this case one is interested in details of
{\it hadron} $\Longleftrightarrow$ {\it QGP} phase transition,
which can only be investigated in this way).

However, one should always keep in mind another possibility,
namely that such behavior can just be due to the fact that, when
presenting our data, out of many particles produced (tens or
thousands at present, say $N$ to be specific) we select only {\it
one} to make the corresponding plots. Single particle
distributions averaged over many events are what is usually
published. But this means that the remaining $(N-1)$ particles
will act as a kind of {\it heath bath}. Assuming that this heath
bath is homogenous and large it is natural to expect that its
action can be described by a single parameter, which we then call
temperature, $T$, and identify with the temperature encountered in
statistical models \cite{LVH}. What one apparently observes then
is just the usual Boltzmann-Gibbs (BG) statistics at work.

The above reasoning assumes that any dynamics of the set of
remaining $(N - 1)$ particles is mostly averaged and what results
looks very much like a state of hadronic matter remaining in {\it
thermal equilibrium} characterized by a temperature $T$. Some
effects, however, can survive this {\it equilibration} process and
can show up as apparent departures from the assumed thermal
equilibrium. This is usually regarded as a departure from BG
statistics and, finally, considered as a kind of failure of the
simple statistical approach. Such observations are therefore a
subject of separate investigations in which many different
(dynamical) ideas compete (like, for example, the production of
resonances and the flow of matter, to name only two)
\cite{dynamics}. However, because there is only a limited amount
of available data, usually one cannot decide which of the proposed
dynamical remedies is right or, if we agree that they should all
be present, in what proportion they show up \footnote{This is
connected with the important problem of how much information given
measurements are providing us with. This is, so far, only
sporadically discussed in the domain of high energy multiparticle
production processes using information theory approach in which
the entropies mentioned here are regarded as measures of
information \cite{Info,Info_new}. This subject is, however, beyond
the scope of our review.}

On the other hand, one can argue that, perhaps, it is the form of
statistical model used which should be modified in such a way as
to account (at least to some extent) for detected irregularities
(i.e., for departures from the BG approach). Therefore, instead of
inventing and investigating different dynamical assumptions, one
can instead investigate the possibility of replacing the usual
statistical model based on BG entropy, $S_{BG}$, by its modified
version based on some other form of entropy discussed in the
literature \cite{Measures}. Such models are widely known nowadays
from other branches of physics and are used whenever a physical
system under investigation shows memory effects of any kind,
experiences long range correlations (i.e., is in a sense "small"
because its size is comparable with the range of forces acting in
it), experiences some intrinsic fluctuations, or the phase space
in which it operates is limited or has fractal structure. In this
review we shall discuss the application of such a model taken in
the form proposed by Tsallis, i.e., in the form based on the so
called Tsallis entropy, $S_T=S_q$ \cite{Tsallis}:
\begin{equation}
S_q = \frac{\left( 1 - \sum_i p_i^q \right)}{q -1}
        \quad \stackrel{q \rightarrow 1}{\Longrightarrow} \quad
        S_{BG} = - \sum_i p_i\, \ln p_i .
 \label{eq:Def}
\end{equation}
Notice that $S_{BG}=S_{q=1}$. The $S_q$ is nonextensive because
for any two independent systems $A$ and $B$ (in the usual sense,
i.e., for which $p_{ij}(A+B)=p_i(A)p_j(B)$), one observes that
\begin{equation}
S_q(A+B) = S_q(A) + S_q(B) + (1-q) S_q(A)S_q(B) ,
         \label{eq:Nonex}
\end{equation}
In this sense the entropic index $q$ is a measure of the
nonextensivity in the system without, however, directly showing up
its cause. This must be provided from elsewhere.

To shed more light on the physics involved here,  let us come back
to the previous reasoning with some effective thermal bath being
formed by the $(N-1)$ particles remaining after selection of the
one used for making final histograms. Notice that, in high energy
multiparticle production reactions observed in  elementary and
heavy-ion collisions we are interested in here, such thermal baths
usually (i.e., after more detailed scrutiny) do not satisfy
conditions allowing us to introduce the notion of thermal
equilibrium in the BG sense: they are always finite and can be
hardly considered as being homogenous (in fact, in many cases they
occupy only a fraction of the allowed phase space
\cite{Correlations} or even have a fractal-like structure
\cite{Fractal} and it is known that usually the hadronizing system
under consideration experiences long range correlations). This
means therefore that such a heath bath cannot be described by a
single parameter $T$. The simplest thing to do seems to be to
allow for some fluctuations of the parameter $T$ and to replace it
by its mean value $T \rightarrow T_0 = \langle T\rangle$ and by
one more parameter describing its fluctuations, using, for
example, the (normalized) variance
\cite{q_interpretation,q_interpretation_1}:
\begin{equation}
\omega = \frac{\langle \left( \frac{1}{T}\right)^2\rangle -
\langle \frac{1}{T} \rangle ^2} {\langle \frac{1}{T} \rangle ^2}.
\label{eq:omega}
\end{equation}
In this approach
\begin{equation}
q\, =\, 1\, +\, \omega . \label{eq:omega1}
\end{equation}
It can be next shown that such a heath bath leads in a natural way
to the following $q$-exponential distribution (called also Tsallis
distributions)\footnote{Except for \cite{q_interpretation_1}
discussion concerning the meaning of the parameter $q$ was limited
to the case of $q>1$ only. As already mentioned in
\cite{q_interpretation_1}, the case of $q<1$ seems not so much
connected with any genuine fluctuations but rather, in some way,
with limitations of the allowed phase space \cite{NUWWx}. It is
worth mention that the idea of possible fluctuations of otherwise
intensive quantities has already been formalized by introducing a
new concept of so called {\it superstatistics} \cite{SuperS}.}:
\begin{equation}
\exp \left( - \frac{X}{\lambda} \right) \Rightarrow \exp_q \left(
- \frac{X}{\lambda}\right) = \left[1 - (1-q)\frac{X}{\lambda}
\right]^{\frac{1}{1-q}} .\label{eq:expqdef}
\end{equation}
This is the power law we were searching for in different reactions
\cite{Info_new,NUWWx,UWWpT,q_approaches,q_approaches1,q_approaches2,NUWW,NUWW1,C_V_Q,Q},
the results of which will be reviewed here\footnote{It should be
noticed that there are also some other investigations on this
subject \cite{qe+e-,ALQ,Wibig,q_approaches_others,AK,BD}, with
rather similar conclusions but which we will not discuss here.}.

The physical picture presented above can be made more formal by
saying that one replaces here the notion of strict {\it local
thermal equilibrium}, customarily assumed in all applications of
statistical models, by the notion of some kind of {\it stationary
state}, which is being formed in the collision and which already
includes some interactions. This concept can be introduced in
different ways. For example, in \cite{SR} it was a random
distortion of energy and momentum conservation caused by the
surrounding system which resulted in the emergence of some
nonextensive equilibrium. In \cite{Biro,BiroK} the two-body energy
composition in transport theory formulation of the collision
process is replaced by a generalized energy sum, $h(E_1,E_2)$,
which is assumed to be associative but which is not necessarily
simple addition and contains contributions stemming from pair
interaction (in the simplest case). It turns out that under quite
general assumptions about the function $h$, a division of the
total energy among free particles is possible. Different forms of
the function $h$ then lead to different forms of entropy formula,
among which one encounters the known Tsallis form. The origin of
this kind of thinking can be traced back to the analysis of the
$q$-Hagedorn model proposed some time ago in \cite{Beck}.

We close this section with a historical note. The recognition that
some, apparently unexpected power law distributions can be due to
fluctuations came to us from the observation in cosmic ray physics
\cite{q_Cosmic} that there exists a {\it long flying component}
(LFC) phenomenon in the propagation of the initial flux of
incoming nucleons. For example, instead of the normally expected
exponential fall off of the depth distribution of the starting
points of cascades, $z$,
\begin{equation}
\frac{dN(z)}{dz} = {\rm const}\cdot \exp\left( -
                z/\lambda\right),  \label{eq:EXP}
\end{equation}
one rather observes an $\exp_q(-z/\lambda)$ distribution, i.e.,
Tsallis like power law behavior given by Eq. (\ref{eq:expqdef})
with $z$ replacing $X$ (here $\lambda \sim 1/\sigma$ is the mean
free path describing propagation of the incoming flux in the
atmosphere with $\sigma$ being the relevant cross section). On the
other hand, some time ago we have shown in \cite{Cosmic} that this
effect can be explained by assuming that hadronic cross sections
should be regarded as fluctuating quantities with widths (defined
as normalized dispersion), $\omega_{\sigma} = \langle
\sigma^2\rangle/\langle \sigma \rangle ^2 - 1$ (and growing
logaritmically with energy). We were at that time prompted by the
fact that such an idea was widely investigated in the usual
hadronic collisions \cite{sigmafluct}\footnote{It is interesting
to notice that this idea has been revitalized very recently in
\cite{newsigmafluct} and connected with the fluctuations in the
gluonic content of hadrons.}. It was then quite natural to connect
the parameter $q$ with fluctuations. This done in
\cite{q_interpretation,q_interpretation_1}, as mentioned above.

We shall review our results in this field in the next Section.
Section \ref{sec:further...} contains some new recent developments
in this field. The final Section contains our conclusions and a
summary.

\section{Review of fluctuations in multiparticle production processes}
\label{sec:review}

High energy collisions result in a multitude of particles of
different kinds being produced. Most are just mesons of all kinds
(overwhelmingly pions). For those who are looking for some new
and/or rare phenomena they form unwanted background which must
somehow be substracted, for others they are a subject of thorough
investigations allowing us to look inside the very early stages of
the collision process as well as at the hadronization stage of the
matter produced (proceeding probably via the formation of the QGP,
for example). In both cases a simple and trustworthy
representation of data is very important, this justifies our
investigations in this field to be reported here.

\subsection{Generalized heat bath - fluctuations of temperature}
\label{GenHB}

We first recall the physical picture behind the generalized heat
bath introduced in Section \ref{sec:intro} which we have proposed
in \cite{q_interpretation,q_interpretation_1}. Our reasoning was
as follows. Suppose we have a thermodynamic system, in a small
(mentally separated) parts of which the temperature can take
different values, i.e., in the whole system it fluctuates with
$\Delta T \sim T$. Let $\xi(t)$ describes stochastic changes of
temperature in time. If the mean temperature of the system
temperature is $\langle T\rangle = T_0$ then, as a result of
fluctuations in some small selected region, the actual temperature
$T'$ equals
\begin{equation}
T'\, =\, T_0\, -\, b\, \xi(t)\, T  ,\label{eq:TTT}
\end{equation}
where the constant $b$ is defined by the actual definition of the
stochastic process under consideration, i.e., by $\xi(t)$, which
is assumed to satisfy the condition that
\begin{equation} \langle
\xi(t) \rangle\, =\, 0 \label{eq:EM}
\end{equation}
and which correlator, $\langle \xi(t)\, \xi(t + \Delta t)
\rangle$, for sufficiently fast changes is equal to
\begin{equation}
\langle \xi(t)\, \xi(t + \Delta t) \rangle\, =\, 2\, D\,
\delta(\Delta t) .\label{eq:COR}
\end{equation}
The inevitable exchange of heat between any selected region of the
system and the rest leads to equilibration of the temperature in
the whole system. The corresponding process of heat conductance is
described by the following equation \cite{LLH},
\begin{equation}
c_p\, \rho\, \frac{\partial T}{\partial t}\, -\, a\, (T'\, -\,
T)\, =\, 0 , \label{eq:HC}
\end{equation}
where $c_p,~\rho$ and $a$ are, respectively, the specific heat
under constant pressure, density and the coefficient of external
conductance. Using $T'$ as defined in (\ref{eq:TTT}) we finally
get the linear differential equation  for the temperature $T$ with
$\tau = b = \frac{c_p\rho}{a}$:
\begin{eqnarray}
\frac{\partial T}{\partial t} + \left[ \frac{1}{\tau} +
\xi(t)\right] T &=& \frac{1}{\tau} T_0 . \label{eq:T}
\end{eqnarray}
It can be now shown that this equation leads to the Lange\-vin
equation with multiplicative noise term  resulting in fluctuations
of the temperature $T$ given in the form of a gamma function
\cite{q_interpretation}
\begin{equation} f(T)\, =\,
\frac{1}{\Gamma(\alpha)}\, \mu\,
 \left(\frac{\mu}{T}\right)^{\alpha-1}\, \exp\left( -\,
\frac{\mu}{T} \right)  \label{eq:FRES}
\end{equation}
 and characterized by the parameters $\mu$ and $\alpha$,
\begin{equation}
\mu = \frac{\phi}{D} \qquad {\rm and} \qquad \alpha =
\frac{1}{q-1} = \frac{1}{\tau\, D} . \label{eq:PAR}
\end{equation}
This is to be compared with Eq. (\ref{eq:omega}) in which now
$\omega = \tau D$. Function $f(T)$ as given by Eq. (\ref{eq:FRES})
is the distribution that should be used to smear the parameter $T$
in the usual exponential distribution of the BG statistical model
and which results in the Tsallis distribution, Eq.
(\ref{eq:expqdef}). To summarize: a small addition of the
multiplicative noise described by a damping constant in the
Langevin equation results in a stationary distribution of particle
momenta, which develops a power-law tail at high
values\footnote{Actually, as shown in \cite{SuperS} when
discussing {\it superstatistics}, there is a whole class of
functions leading from $\exp(X)$ to $\exp_q(X)$. But only this has
a simple physical interpretation as presented here. More general
version of Langevin equation containing also additive noise have
been considered in \cite{BiroJ}.}.

\subsection{Transverse and longitudinal dynamics}
\label{sec:TL}

To begin our presentation we first set the stage. The
characteristic pattern of the multiparticle production processes
is that most of the secondaries are produced with small transverse
momenta $p_T$ (mostly below $1$ GeV) and are therefore
concentrated in the longitudinal phase space given by the
longitudinal momenta $p_L$ (which is described in terms of the
rapidity $y=\frac{1}{2}\ln \frac{E+p_L}{E-p_L}$, where $E =
\sqrt{m_T^2 + p_L^2}$ with $m_T = \sqrt{m^2 + p_T^2}$ being the so
called transverse mass ($m$ is mass of the particle); in other
notation $E = m_T \cosh y$ and $p_L = m_T \sinh y$). The terms
{\it transverse} and {\it longitudinal} are defined with respect
to the direction of the colliding particles. Data are presented as
distributions either in $p_T$ or in $y$. In both cases they show
exponential behavior either in $p_T$ or in the energy $E = \langle
m_T\rangle \cosh y$ (with $\langle m_T\rangle =\sqrt{m^2 +\langle
p_T\rangle^2}$):
\begin{equation}
\frac{dN}{dp_T}\! =\! C_{p_T}p_T \exp\left( -
\frac{p_T}{T}\right);\quad \frac{dN}{dy}\! =\! C_y \exp\left( -
\frac{E}{T}\right). \label{eq:pTE}
\end{equation}
One observes dramatic differences in both distributions reflected
by the differences in the values of the parameter $T$, which is of
the order of one hundred MeV in $p_T$ space (where $T=T_{p_T}$ and
is universal, i.e., essentially energy independent) and tens of
GeV (depending on the energy of collision) in $p_L$ (or $y$) space
(where $T=T_{p_L}$ and depends on energy). This means that the two
distributions reflect different physics: those in $p_T$ space are
believed to be essentially "thermal-like" and subject to a
thermodynamic interpretation whereas, those in $p_L$ space are
sensitive to the available energy and to the multiplicity of
produced secondaries. Because of this their fluctuation patterns
will be different, i.e., when described by Tsallis power-like form
Eq. (\ref{eq:expqdef}) the corresponding parameters $(q_T - 1)$
and $(q_L - 1)$ will differ dramatically. Also the physical
meaning of these parameters will be different reflecting different
sources of fluctuations.

\subsection{Longitudinal phase space}
\label{sec:Lps}

\begin{figure}[h]
\begin{center}
\resizebox{0.4\textwidth}{!}{
  \includegraphics{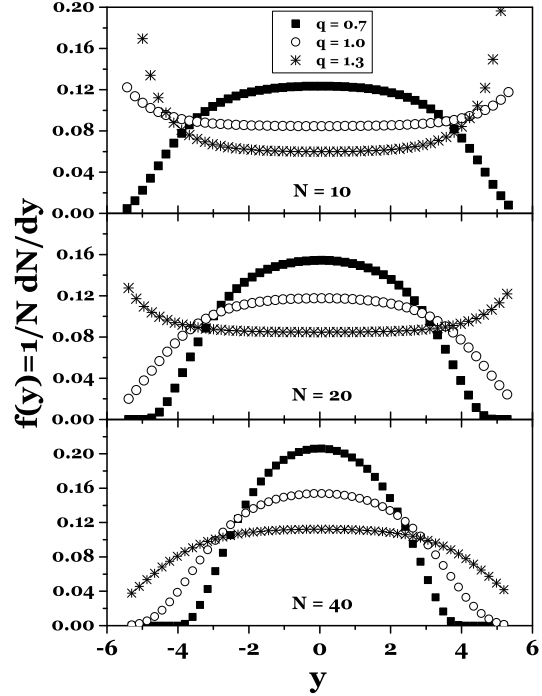}
  }
\caption{Examples of the most probable rapidity distributions as
given by Eq. (\ref{eq:pTE})  for hadronizing mass $M = 100$ GeV
decaying into N secondaries of (transverse) mass $m_T = 0.4$ GeV
each for different values of parameter $q=Q_L$ (reproduced by
permission of Springer-Verlag from \cite{q_approaches2}).}
\label{FigL}
\end{center}
\vspace{-0.5cm}
\end{figure}

We start with the longitudinal distribution in rapidity (averaged
over $p_T$). In Fig. \ref{FigL} one observes that $q<1$ is
effectively cutting off the allowed longitudinal phase space (here
defined by the initial available energy $M = 100$ GeV and assumed
constant transverse mass $m_T=0.44$ GeV and weakly depending on
the assumed multiplicity of the produced particles $N$). Actually,
from Eq. (\ref{eq:expqdef}) it is obvious that only such
combinations of $q$ and $X$ and $\lambda$ are allowed for which
$[1-(1-q)X/\lambda] > 0$. In \cite{NUWWx}, when fitting
longitudinal distributions without restricting the available
energy by introducing the so called inelasticity coefficient $K
<1$, the only role of $q$, which was found to be $q<1$ there, was
to limit the amount of energy used (showing the necessity of
introducing inelasticity when considering multiparticle production
processes, cf., \cite{inel} for review on this subject). For $q>1$
one observes a visible enhancement of distribution tails.

\begin{figure}[h]
\begin{center}
\resizebox{0.45\textwidth}{!}{
   \includegraphics{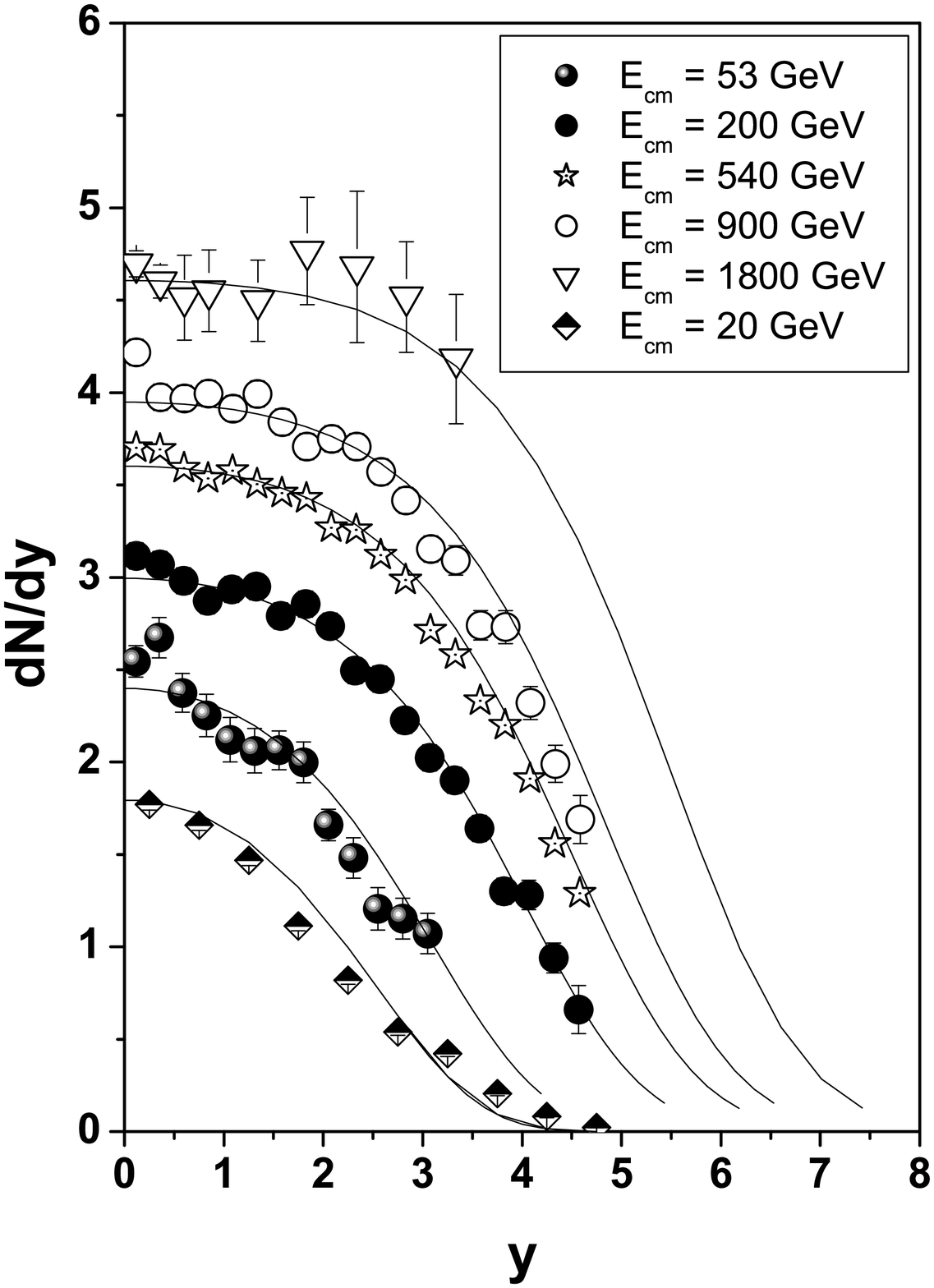}
   \includegraphics{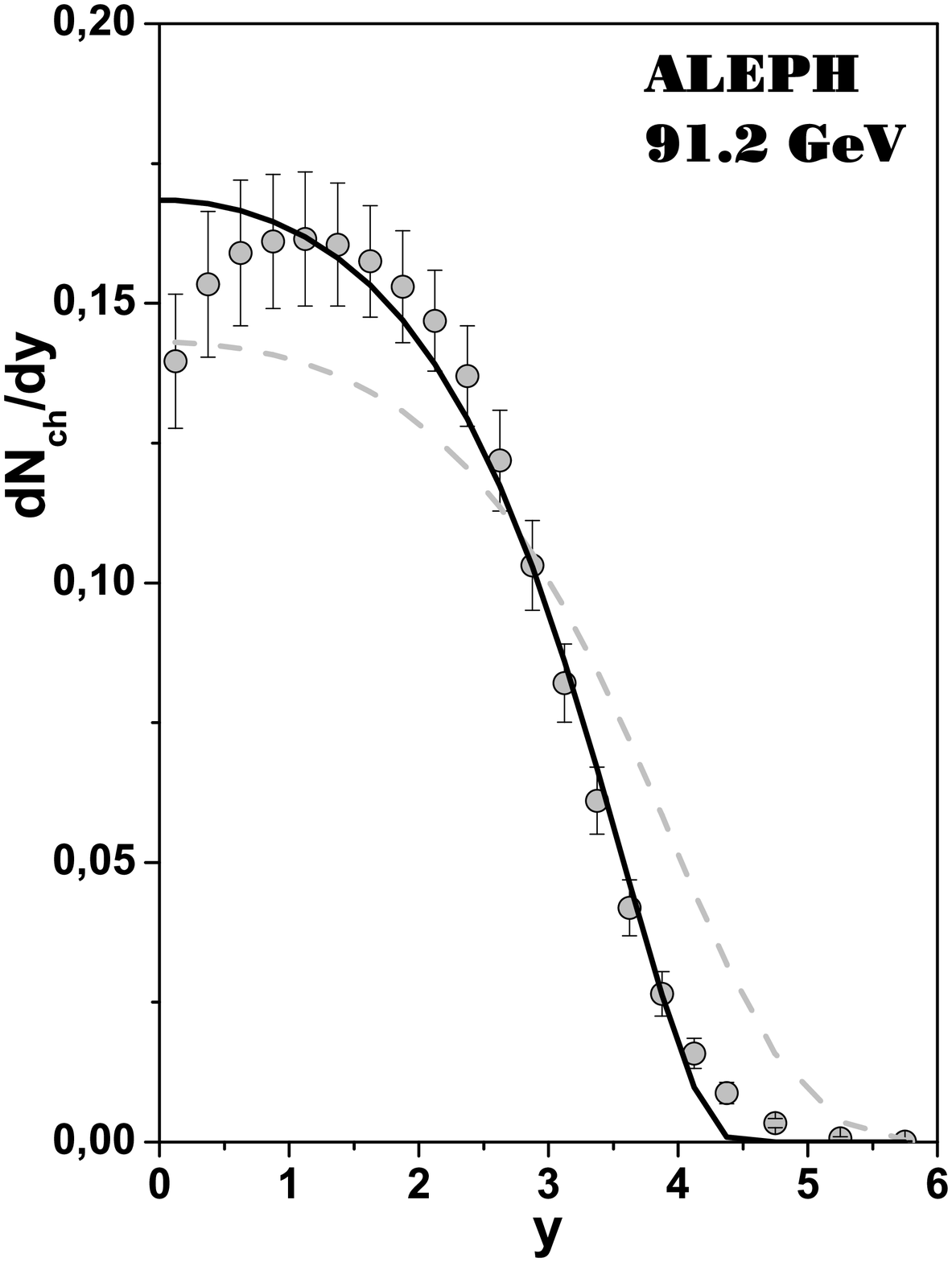}
} 
\caption{Examples of applying the nonextensive approach to
longitudinal distributions. Left panel: fit to rapidity spectra
for charged pions produced in $pp$ and $\bar{p}p$ collisions at
different energies \cite{Datapp1}. Right panel: rapidity spectra
measured in $e^+e^-$ annihilations at $91.2$ GeV \cite{Data5}
(dotted line is for $K_q=1$ and $q=1$ whereas the full line is our
fit with $K_q=1$ and $q=0.6$). (Reprinted from Physica A344, F.S.
Navarra, O.V. Utyuzh, G. Wilk and Z. W\l odarczyk, "Information
theory in high-energy physics (extensive and nonextensive
approach)", 568, Copyright (2004), with permission from Elsevier;
http://www.elsevier.com.). } \label{FigR1}
\end{center}
\vspace{-5mm}
\end{figure}
 Physica A
                {\bf 344}, 568  (2004)

\begin{figure}[h]
\vspace{-2mm}
\begin{center}
\resizebox{0.45\textwidth}{!}{
   \includegraphics{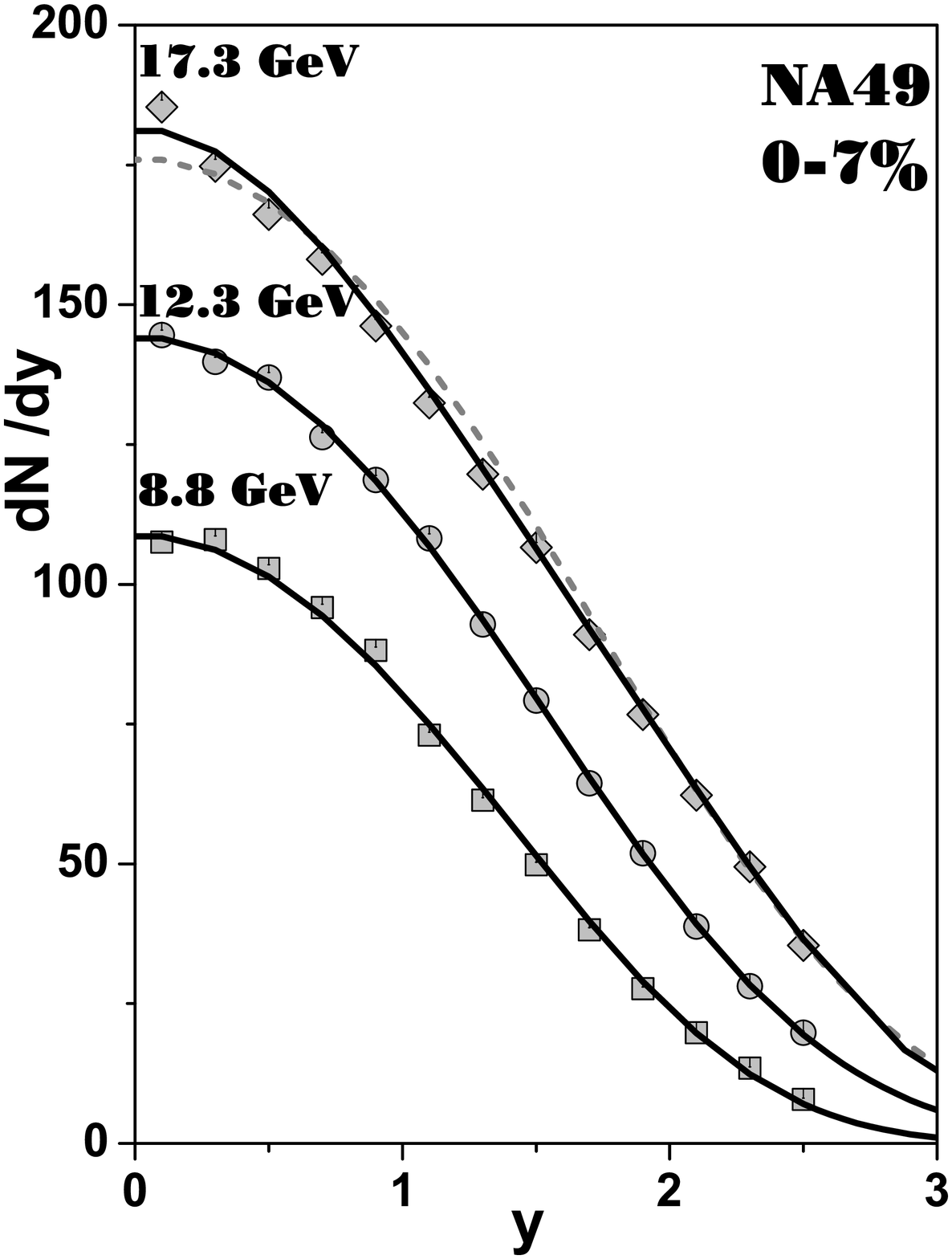}
   \includegraphics{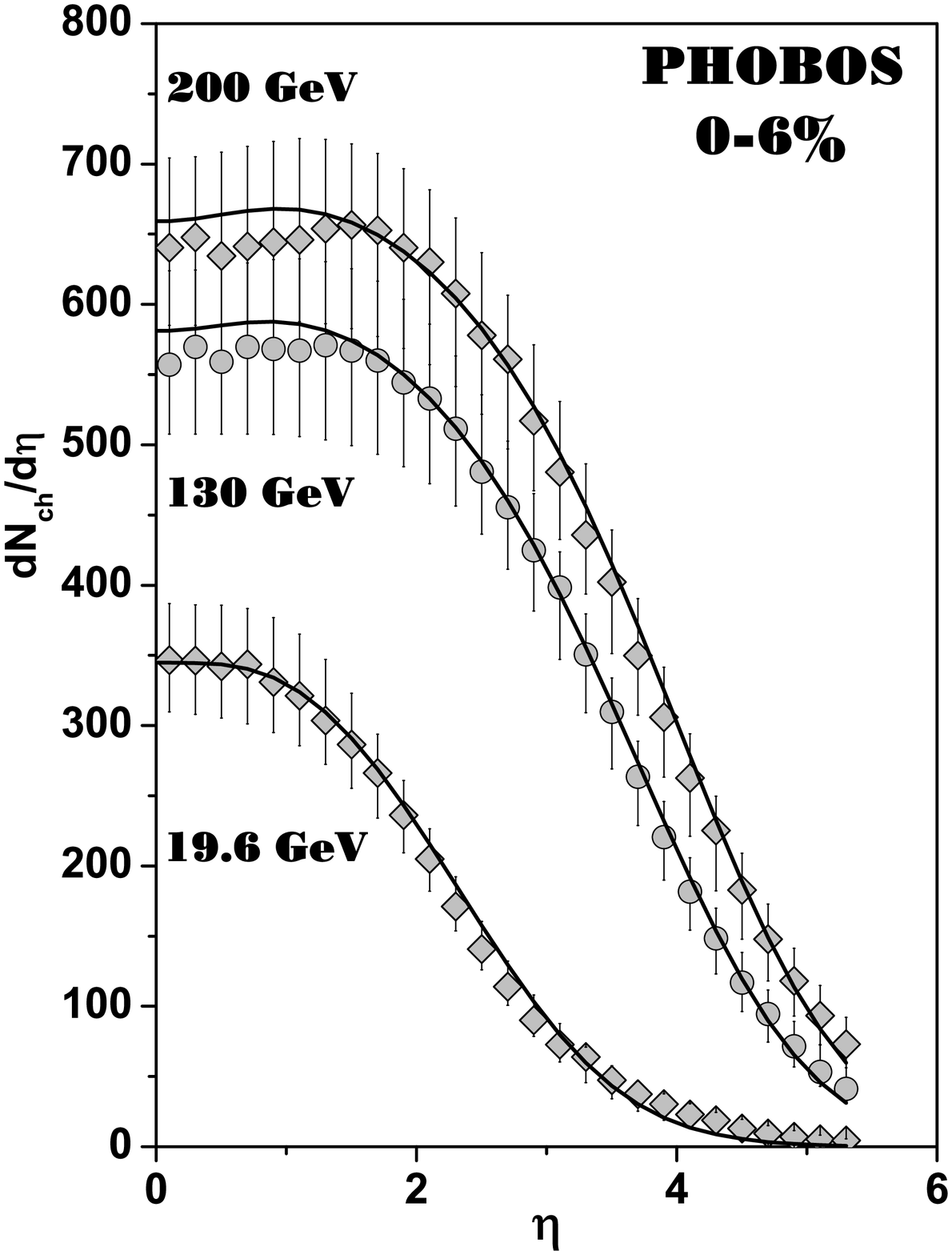}
} 
\caption{Examples of applying a nonextensive approach to
longitudinal distributions. Left panel: fits to NA49 data for
Pb+Pb collisions \cite{DataAA}. Right panel: fits to PHOBOS data
for Au+Au collisions \cite{Data4} . (Reprinted from Physica A340,
F.S. Navarra, O.V. Utyuzh, G. Wilk and Z. W\l odarczyk,
"Information theory approach (extensive and nonextensive) to
high-energy multiparticle production processes", 467, Copyright
(2004), left panel, and from Physica A344, F.S. Navarra, O.V.
Utyuzh, G. Wilk and Z. W\l odarczyk, "Information theory in
high-energy physics (extensive and nonextensive approach)", 568,
Copyright (2004), right panel; both with permission from Elsevier;
http://www.elsevier.com.).} \label{FigR2}
\end{center}
\vspace{-6mm}
\end{figure}

The examples of fits to the actually observed single particle
distributions in rapidity are shown in Figs. \ref{FigR1} and
\ref{FigR2}. In the left panel of Fig. \ref{FigR1} (see
\cite{NUWW} for details) results for $pp$ and $p\bar{p}$
collisions at energies varying between $\sqrt{s}=20$ GeV to $1800$
GeV are displayed. From each listed energy of collision,
$E_{cm}=\sqrt{s}$, only a fraction $K_q$ has been used for the
production of secondaries (according to \cite{inel}). The other
input was the mean multiplicity of charged secondaries produced in
nonsingle diffractive reactions at given energy: $\bar{n}_{ch} =
-7.0 + 7.2 s^{0.127}$ \cite{Thermodynamics} (corresponding to the
total number of produced particles, $N=\frac{3}{2}\bar{n}_{ch}$).
The allowed phase space is one dimensional with only a small
energy dependence of the mean transverse momentum allowed,
$\langle p_T\rangle = 0.3 + 0.044\ln\left(\sqrt{s}/20\right)$
\cite{Thermodynamics} (all secondaries will be assumed to be pions
of mass $\mu = 0.14$ GeV).

\begin{figure}[h]
\begin{center}
\resizebox{0.4\textwidth}{!}{
   \includegraphics{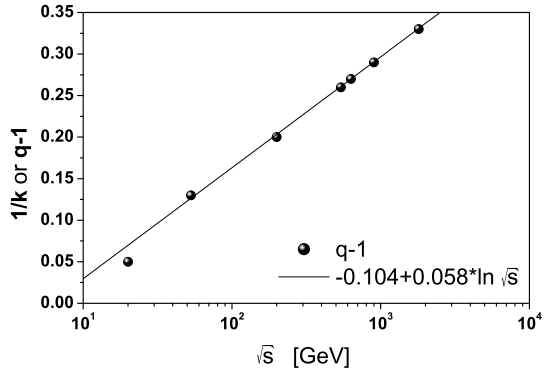}
} 
\caption{The values of the nonextensivity parameter $q$ obtained
in fits shown in the left panel of Fig. \ref{FigR1} compared with
the values of the parameter $k$ of a Negative Binomial
distribution fit to the corresponding multiplicity distributions
as given in \cite{Thermodynamics}. ( Reprinted Fig. 6 with
permission from F.S. Navarra, O.V. Utyuzh, G. Wilk and Z. W\l
odarczyk, Phys. Rev. D {\bf 67}, 114002 (2003). Copyright (2003)
by the American Physical Society;
URL:http://link.aps.org/abstract/PRD/v67/e114002; DOI: 10.1103/
PhysRevD.67.114002.).} \label{FigQ}
\end{center}
\vspace{-5mm}
\end{figure}

As discussed in detail in \cite{NUWW}, one gets in this case not
only the parameter $q$ but also the true inelasticity, $K$, of the
reaction (in fact, even, for the first time, its distribution,
$\chi(K)$). Let us, however, concentrate on the parameter $q$,
which bears information on fluctuations. It turns out to be energy
dependent as presented in Fig. \ref{FigQ}. Surprisingly enough it
turned out that the same behavior is observed for the inverse of
$k$ characterizing the so called  Negative Binomial distribution
(NB) \cite{Thermodynamics} of the multiplicity of observed
secondaries, which depends on two parameters: the mean
multiplicity $\langle n_{ch}\rangle$ and the parameter $k$ ($k\ge
1$) affecting its width ($\sigma)n_{ch}$ is dispersion),
\begin{equation}
\frac{1}{k}\, =\, \frac{\sigma^2(n_{ch})}{\langle
n_{ch}\rangle^2}\, -\, \frac{1}{\langle n_{ch}\rangle}.
\label{eq:k}
\end{equation}
For $k\rightarrow 1$ NB approaches a geometrical distribution
where\-as for $k^{-1}\rightarrow 0$ it approaches a Poissonian
distribution. In general it is found \cite{Thermodynamics} that $
\frac{1}{k} = -0.104 + 0.058\cdot \ln \sqrt{s}$, which fits the
obtained values very nicely, cf. Fig. \ref{FigQ}.

To fully understand the possible physical meaning of the parameter
$q (=q_L)$ in this case let us remind ourselves that, in general,
the nonextensivity parameter $q$ summarizes the action of several
factors, each of which leads to a deviation from the simple form
of the extensive BG statistics, as was mentioned before, out of
which we are interested most in the possible intrinsic
fluctuations existing in the hadronizing system
\cite{q_interpretation}. Notice that in our fits we have not
explicitly accounted for the fact that each event has its own
multiplicity, $N$, but we have used only its mean value, $\langle
N\rangle$, as given by experiment where $\langle N\rangle = \sum N
P(N)$ with $P(N)$ being the multiplicity distribution
\footnote{Actually, we have used only its charged part, $\langle
n_{ch}\rangle$, assuming that $N = \frac{3}{2}\langle
n_{ch}\rangle$, i.e., neglecting in addition also possible
fluctuations between the number of charged and neutral
secondaries.}. On the other hand, the parameter $T$ in this case
is not so much a temperature, but only a kind of "partition
temperature", understood as mean energy per produced particle,
i.e., $T \sim W/\langle N\rangle$ (where $W=K\sqrt{s}$, where
$\sqrt{s}$ is the total energy of collision) \cite{partitionT}.
Therefore in this case one can just as well speak about the
fluctuations of $\langle N\rangle$. Following therefore the ideas
of \cite{q_interpretation} we would like to draw attention to the
fact that the value of $k^{-1}$ may also be understood as the
measure of fluctuations of the mean multiplicity (for example, in
the usual Poissonian multiplicity distribution characterized just
by a single parameter,  the constant mean multiplicity $\bar{n}$),
and in the case when such fluctuations are given by a gamma
distribution with normalized variance $D(\bar{n})$, one obtains
the Negative Binomial multiplicity distribution with
\begin{equation}
\frac{1}{k}\, =\, D(\bar{n})\, =\,
\frac{\sigma^2\left(\bar{n}\right)}{\langle \bar{n}\rangle^2} .
\label{eq:D}
\end{equation}
This is because in this case one has \cite{Shih}:
\begin{eqnarray}
P(n) &=& \int_0^{\infty} d\bar{n}
\frac{e^{-\bar{n}}\bar{n}^n}{n!}\cdot
         \frac{\gamma^k \bar{n}^{k-1} e^{-\gamma \bar{n}}}{\Gamma (k)}\nonumber\\
         &=&
   \frac{\Gamma(k+n)}{\Gamma (1+n) \Gamma (k)}\cdot
   \frac{\gamma^k}{(\gamma +1)^{k+n}}, \label{eq:PNBD}
\end{eqnarray}
where $\gamma = \frac{k}{\langle \bar{n}\rangle}$.

Therefore the situation in longitudinal phase space is following:
When there are only statistical fluctuations in the hadronizing
system one expects a Poissonian form of the corresponding
multiplicity distributions. The existence of intrinsic (dynamical)
fluctuations means that one allows the mean multiplicity $\bar{n}$
to fluctuate. It is natural to assume that these fluctuations
contribute predominantly  to the longitudinal phase space, i.e.,
that $D(\bar{n}) = q - 1$ and that
\begin{equation}
q\, =\, 1\, +\, \frac{1}{k}. \label{eq:qk}
\end{equation}
This is observed in the data.

The right hand panel of Fig. \ref{FigR1} displays results for
$e^+e^-$ annihilations for which, by definition, $K_q = 1$
(because always all the energy of initial leptons is available for
the production of secondaries) and which can  be fitted {\it only}
with $q<1$ (in our case $q=0.6$). This should be contrasted with
results obtained describing the $p_T$ distributions instead where
one finds $q>1$ \cite{qe+e-}. This point deserves closer scrutiny.
The result for $q=1$ clearly shows that observed discrepancies are
not connected with the particular value of $q$, but rather with
some additional mechanisms operating here, the action of which
would, however, change our results only slightly (for example, a
possibility of two rather than one source or $y$-dependent
$\langle p_T\rangle$, as mentioned already in \cite{Info_new}).
With the above reservations, let us then take a closer look at the
possible origin of $q < 1$. We have already encountered a similar
situation when in \cite{NUWWx} $q < 1$ was simply closing the
allowed a priori phase space, acting therefore as inelasticity
parameter $K$. When considered as a signal of fluctuations
(similar to the $q>1$ case) \cite{q_interpretation_1} it causes
trouble because in this case the temperature $T$ does not reach an
equilibrium state, in fact one now has that the source term (right
hand side of Eq. (\ref{eq:T})) is $T_0/\tau - (q-1)E/\tau$ rather
than $T_0/\tau$ used for for $q > 1$ case (cf.,
\cite{q_interpretation_1}). This means than that in this case we
have a kind of dissipative transfer of energy from the region
where (due to fluctuations) the temperature $T$ is higher. It
could be any kind of convection-type flow of energy; for example,
it could be connected with the emission of particles from this
region (for example, in our case from a quark ($q$) and antiquark
($\bar{q}$) jets formed in the first $e^+e^- \rightarrow q +
\bar{q}$ to gluons and $q\bar{q}$ pairs and later on to finally
observed hadrons). This means that $q<1$ signals that in the
reaction considered, where $K_q=1$ and where we have to account
for the whole energy exactly, conservation laws start to be
important and there is no possibility for a stationary state with
constant final temperature to develop. Instead, the  temperature
$T$ depends on the energy\footnote{Actually, in the case
considered in \cite{q_interpretation_1} fluctuations depend on
energy  in the same way leaving the relative variance $\omega$
constant and leading to $ q = 1-\omega$.}, and for large energies
tends to zero (notice that in this case one has a limitation on
the allowed energy of the produced secondaries: $E \le
T_0/(1-q)$). This is not the case for the $p_T$ distribution
analysis \cite{qe+e-} because most $p_T$ are small and are not
influenced by the conservation laws but instead reflect a kind of
stationary state with $q>1$.

Now look at the left hand panel of Fig. \ref{FigR2}. It shows fits
to NA49 data \cite{DataAA} on $\pi^-$ production in $PbPb$
collisions at three different energies per nucleon. The obtained
values of nonextensivities and the corresponding inelasticities,
$(q;K_q=3\cdot K^{\pi^-}_q)$ are: $(1.2;0.33)$ for $17.3$ GeV,
$(1.164;0.3)$ for $12.3$ GeV and $(1.04;0.22)$ for $8.6$ GeV. The
origin of $q>1$ in this case is not yet clear. The inelasticity
seems to grow with energy. It is also obvious that, for higher
energies, some new mechanism starts to operate because we cannot
obtain agreement with data using only energy conservation. The
best fit for $17.3$ GeV for NA49 data actually for the case of
$q=1$ and two sources separated in rapidity by $\Delta y = 0.83$
(cf., \cite{Info_new} for other details).

Finally, the right hand panel of Fig. \ref{FigR2} presents fits to
pion production in Au+Au collisions \cite{Data4} for the most
central events (covering collisions proceeding with impact
parameter range $0-6\%$) \footnote{Actually these data are
originally presented not for the rapidity $y$ defined by the
energy $E$ and the the longitudinal momentum $p_L$ but for the so
called pseudorapidity $\eta$ defined by the total momentum $p$ and
the longitudinal momentum $p_L$ instead. There is therefore some
ambiguity when transferring them from $\eta$ to $y$ because of the
pour knowledge of the rapidity dependence of the mean transverse
momentum needed for such operation.}. They can be fitted by
choosing $K_q=1$ and then $q= 1.29$, $1.26$ and $1.27$ for
energies $19.6$, $130$ and $200$ GeV, respectively (cf.,
\cite{NUWW1} for other details). As before, the origin of $q>1$ in
this case is not yet clear.

Although the situation in AA collisions is not yet clear, we are
quite confident that interpretation of the $q$ parameter offered
here remains valid. But, before settling this, one point has to be
addressed. Namely, the above $q$ were in fact $q_L$ responsible
for the longitudinal dynamics only. On the other hand,
multiplicity distributions are sensitive to $p=\sqrt{p^2_L +
p^2_T}$ and, as we have seen here, both $p_L$ and $p_T$ show
traces of fluctuations by leading to $q>1$. However, as we shall
see below, $(q_T -1) << (q_L -1)$ (what fits nicely the fact that
$p_T$ space is very limited in comparison to $p_L$ one). Because
there are no data measuring $p_T$ distributions at all values of
rapidity $y$, i.e., providing correlations between parameters
$(T;q) = (T_L;q_L)$ for longitudinal momenta (rapidity)
distributions and $(T;q) = (T_T;q_T)$ for transverse momenta
distributions, we offer only the following approximate answer.
Noticing that $q - 1 = \sigma^2(T)/T^2$ (i.e., it is given by
fluctuations of total temperature $T$) and assuming that
$\sigma^2(T) = \sigma^2(T_L) + \sigma^2(T_T)$, one can estimate
that the resulting values of $q$ should not be too different from
\begin{equation}
q\, =\, \frac{q_L\, T_L^2\, +\, q_T\, T^2_T}{T^2}\, -\,
        \frac{T^2_L\, +\, T^2_T}{T^2}\, +\, 1,
\label{eq:qqq}
\end{equation}
which, for $T_L \gg T_T$, as is in our case, leads to the result
that $q \sim q_L$, i.e., it is given by the longitudinal
(rapidity) distributions only.

\subsection{Transverse phase space}
\label{sec:Tps}

\begin{figure}[h]
  \begin{center}
   \vspace{2mm}
   \includegraphics[width=3.4cm]{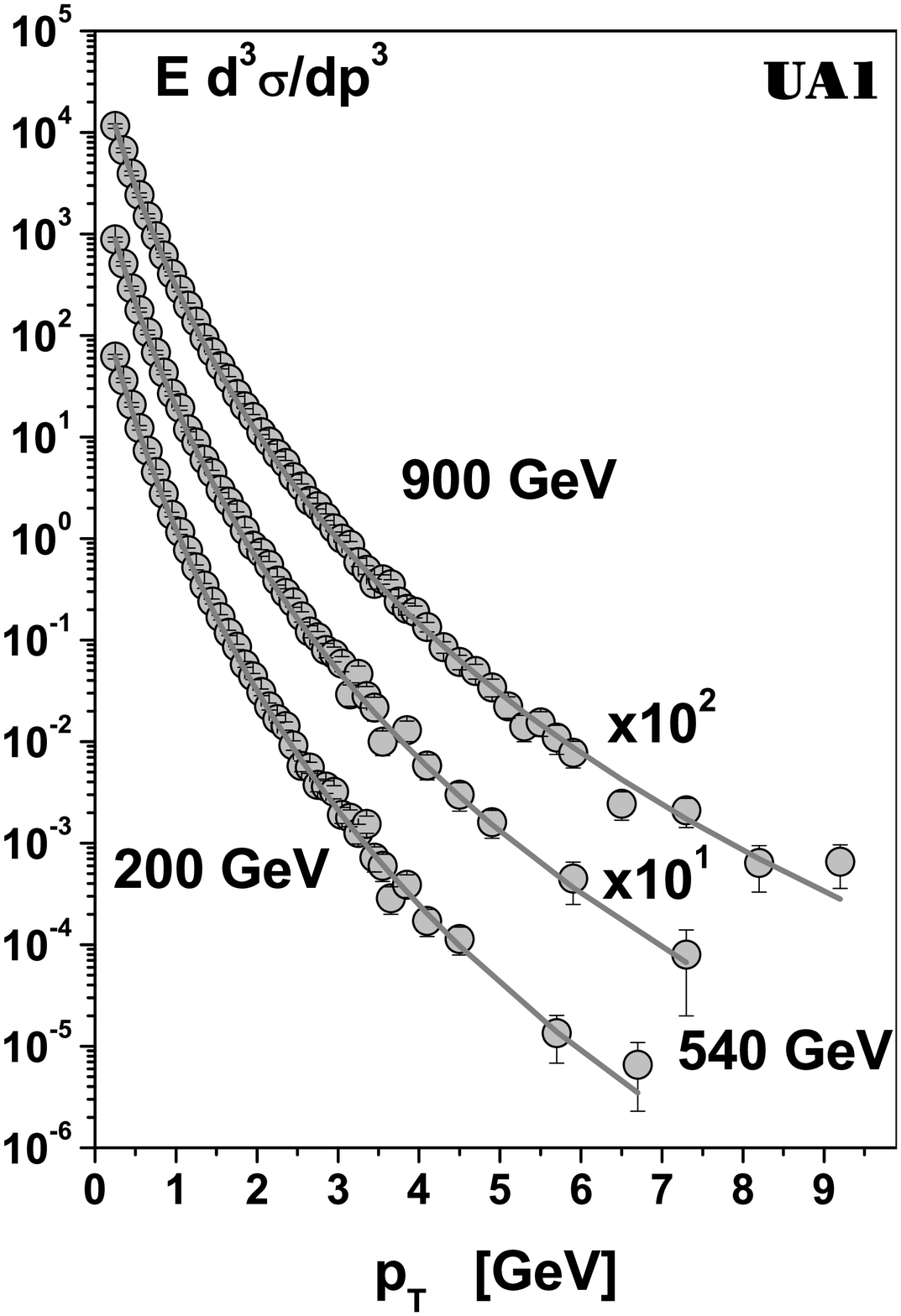}
   \includegraphics[width=4.4cm]{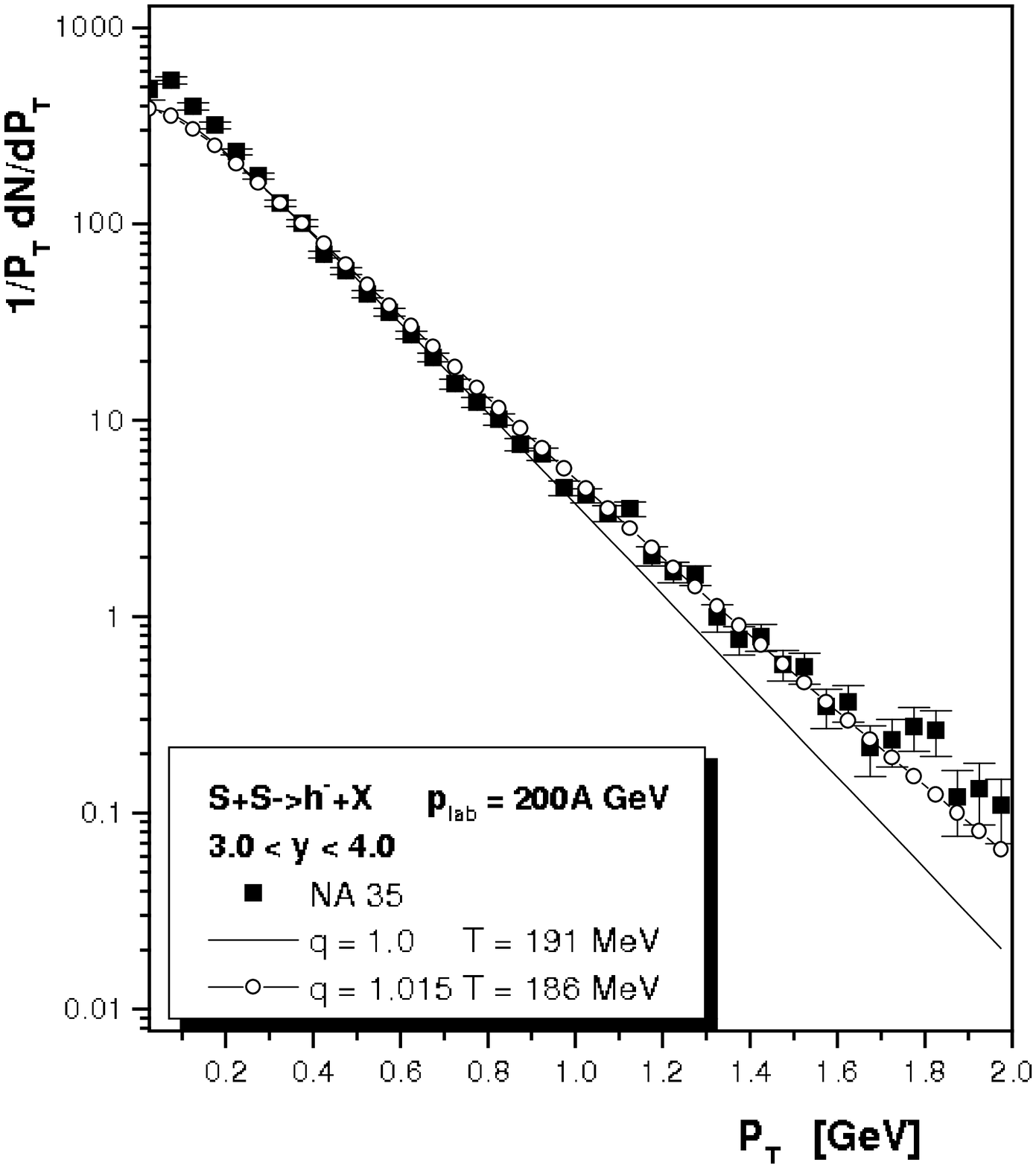}
   \caption{Examples of applying a nonextensive approach to transverse
            momenta distributions. Left panel: fits to $p_T$ spectra
            from $p\bar{p}$ UA1 experiment \cite{Datapt} for different
            energies (see text for details (reprinted from Physica A340,
            F.S. Navarra, O.V. Utyuzh, G. Wilk and Z. W\l odarczyk,
            "Information theory approach (extensive and nonextensive)
            to high-energy multiparticle production processes", 467,
            Copyright (2004), reproduced by permission
            of Elsevier, http://www.elsevier.com.).
            Right panel: fits to $S+S$ data from \cite{DatapT}
            (reproduced by permission of IOP Publishing Ltd from  \cite{UWWpT}). }
   \label{FigpT}
  \end{center}
  \vspace{-5mm}
\end{figure}

As discussed before, transverse phase space seems to be mainly
dominated by the thermodynamical-like effects governed by the
temperature $T$ \cite{Thermodynamics}. It is therefore the best
place too look for any fluctuations of temperature, i.e., to look
for any deviation of the the inverse slope of transverse momenta
distributions, $dN/dp_T$, from an exponential shape. That such
deviations are really observed is seen in Fig. \ref{FigpT}. On the
left hand panel we can see fits to $p_T$ spectra measured by the
UA1 experiment \cite{DatapT} in $p\bar{p}$ at different energies
using Tsallis distribution, Eq. (\ref{eq:expqdef}), with
$X/\lambda \rightarrow p_T/T$ and with the following values of
$(T=T_T$[GeV]$,q=q_T)$: $(0.134,~1.095)$, $(0.135,~ 1.105)$ and
$(0.14,~ 1.11)$ for energies $200$, $500$ and $900$ GeV,
respectively (the values of the parameter $q$ obtained in analysis
of transverse momenta in elementary $e^+e^-$ reaction is similar
\cite{qe+e-}). These values should be compared with the
corresponding values of $(T = T_L; q = q_L)$ previously observed
for rapidity distributions, which are equal to, respectively:
$(11.74~, 1~.2)$, $(20.39,~ 1.26)$ and $(30.79,~ 1.29)$ at
comparable energies, cf. \cite{NUWW}.

The right hand panel of Fig. \ref{FigpT} shows an example of
similar behavior observed for nuclear collisions. Such collisions
are of special interest as they are the only place where a new
state of matter, the Quark Gluon Plasma, can be produced
\cite{QGP} and, because of this, they are intensively investigated
using a nonextensive approach (see, for example,
\cite{q_interpretation_1,C_V_Q,ALQ,Wibig,BD}). As one can see, the
best fit is obtained for $q>1$, albeit in this case the value of
$(q-1)$, which is the real measure of fluctuations, is noticeably
smaller than in the case of elementary reactions mentioned above.
On the other hand, although very small ($|q-1| \sim 0.015$), this
deviation leads to a quite substantial relative fluctuations of
the temperature existing in the nuclear collisions, namely one
gets that $\Delta T/T \simeq 0.12$.

The question then arises: if this is treated seriously, what we
are really measuring, what physical observable does it correspond
to? It is important to stress that these are fluctuations existing
in small parts of a hadronic system in respect to the whole system
rather than of the event-by-event type for which, $$\Delta T/T =
0.06/\sqrt{N} \rightarrow 0$$ for large $N$. The answer is that
the measured fluctuations provide a direct measure of the total
heat capacity $C$ of the system \cite{Capacity},
\begin{equation}
\frac{\sigma^2(\beta)}{\langle \beta\rangle ^2}\, =\,
\frac{1}{C}\, =\, \omega\, =\, q - 1 , \label{eq:C}
\end{equation}
($\beta =\frac{1}{T}$) in terms of $\omega = q - 1$. Therefore,
single particle distributions of produced secondaries, if only
measured very precisely, can {\it a priori} provide us information
not only on the temperature $T$ of the hadronizing system but
also, when investigated using a nonextensive approach, give us
information (via value of $q - 1$) on its total heat capacity $C$.
In this way one can not only check whether some (approximate)
thermodynamical state is formed in a single collision but also
what are its thermodynamical properties - a very important
feature, especially in what concerns the existence and type of the
possible phase transitions \cite{QGP}.

\begin{figure}[h]
 \resizebox{0.45\textwidth}{!}{
  \includegraphics{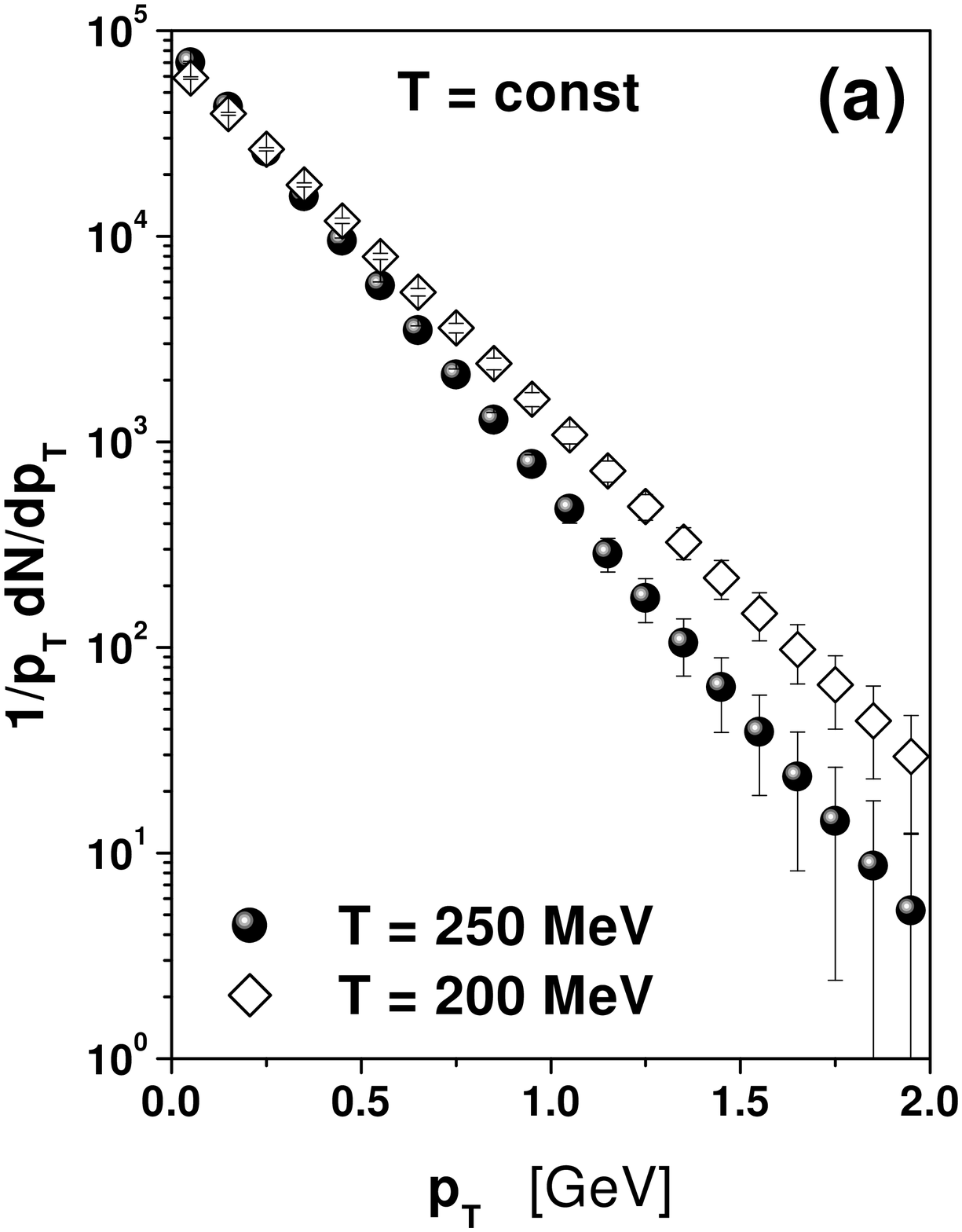}
  \includegraphics{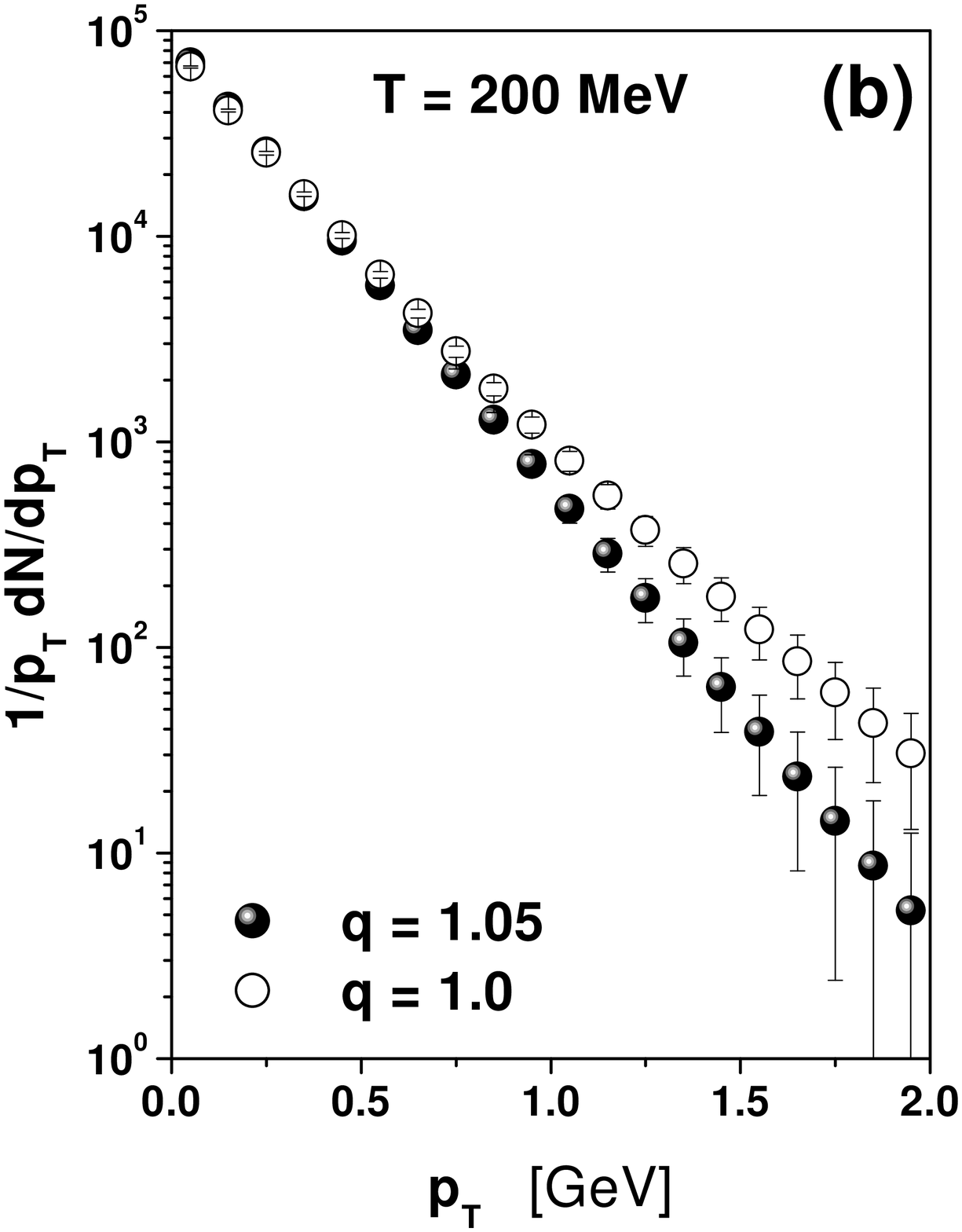}
} 
\caption{$(a)$ Normal exponential $p_T$ distributions
         i.e., $q=1$) for $T = 200$ MeV (black symbols)
        and $T = 250$ MeV open symbols). $(b)$ Typical
        event from central $Pb+Pb$ at $E_{beam}=3~A\cdot$TeV
        for $=200$ MeV for $q=1$ (black symbols) exponential
        dependence and $q=1.05$ (open symbols).
        (Reprinted from Physica A305, G. Wilk and Z. W\l odarczyk,
            "Application of nonextensive statistics to particle
            and nuclear physics", 227, Copyright (2002), reproduced by permission
            of Elsevier, http://www.elsevier.com.). }
        \vspace{-5mm}
        \label{FigTq}
\end{figure}

The next question is: how plausible is such a program? The point
is, as discussed above, that one performs fits using $T$ and $q$
in a Tsallis distributions rather than only $T$ in the usual
exponential ones. However, the corresponding data on $p_T$ are
effectively integrated over the longitudinal phase space (or, at
least a part of it) and are averaged over many events. The best
thing would be to observe such an effect in a single events, then
an event-by-event analysis of data would be possibly. Fig.
\ref{FigTq} shows what we can expect. Two scenarios are
demonstrated there: $(a)$ $T$ is constant in each event but
(because, for example, of different initial conditions) it
fluctuates from event to event and $(b)$ $T$ fluctuates in each
event around some mean value $T_0$. We have chosen for comparison
a typical event obtained in simulations performed for central
$Pb+Pb$ collisions taking place for beam energy equal
$E_{beam}=3~A\cdot$TeV (expected shortly in ALICE experiment at
LHC). Density of particles in the central region (defined by
rapidity window $-1.5 <y< 1.5$) is chosen to be equal to
$\frac{dN}{dy} = 6000$. In case $(a)$ in each event one expects
exponential dependence with $T=T_{event}$ and possible departure
from it would occur only after averaging over all events. It would
reflect fluctuations originating from different initial conditions
for each particular collision. This situation is illustrated in
Fig. \ref{FigTq} (a) where $p_T$ distributions for $ T = 200$ MeV
(black symbols) and $T = 250$ MeV (open symbols) are presented.
Such values of $T$ correspond to typical uncertainties in $T$
expected at LHC accelerator at CERN. Notice that both curves
presented here are straight lines. In case $(b)$ one should
observe departure from the exponential behavior already on the
single event level and it should be fully given by $q>1$. This
reflects a situation when, due to some intrinsically dynamical
reasons, different parts of a given event can have different
temperatures, as we have discussed above. In Fig. \ref{FigTq} (b)
black symbols represent exponential dependence obtained for $T =
200$ MeV (the same as in Fig. \ref{FigTq} (a) ), open symbols show
the power-like dependence as given by (\ref{eq:expqdef}) with the
same $T$ and with $q=1.05$ (notice that the corresponding curve
bends slightly upward here). In this typical event we have $\sim
18000$ secondaries, i.e., practically the maximal possible number.
Notice that here points with highest $p_T$ already correspond to
single particles. As one can see, experimental differentiation
between these two scenarios will be very difficult, although not
totally impossible. On the other hand, if successful it would be
very rewarding \footnote{It is interesting to realize that for the
Planckian gas at $T=186$ MeV, occupying volume of the order of the
volume of sulfur nucleus, one gets $C=34.4$ per degree of freedom,
which leads, using Eq. (\ref{eq:C}), to $q=1.015$ obtained for
such system for the $p_T$ dependence of produced secondaries.}.

The following remarks are worth to be done at this point.
\begin{itemize}
\item We are using rather freely the notion of fluctuating
temperature. The question then arises whether is makes sense. Not
going into a detailed dispute, we would only like to mention at
this point that traces of this idea can already be found in
\cite{GEN,QCD} \footnote{For those interested in discussion on the
problem of internal consistency (or inconsistency) of the notion
of fluctuations of temperature in thermodynamics we refer to
\cite{Capacity,GEN,FLUCT}.}. In particular it is important when
discussing some peculiarities of the phase diagrams, which are
important when addressing the question of possible phase
transitions between QCD and normal matter \cite{QCD}. What we want
to do is to bring to ones attention the fact that event-by-event
analysis allows us (at least {\it in principle}) to detect
fluctuations of temperature taking place {\it in a given event}.
This is more than an indirect measure of fluctuations of $T$
proposed some time ago in \cite{KM} or more direct fluctuations of
$T$ {\it from event to event} discussed in \cite{GEN}. \item As
the heat capacity $C$ is proportional to the volume, $C \propto
V$, in our case V would be the volume of the interaction (or
hadronization), it is expected to grow with volume and,
respectively, $q$ is expected to decrease with $V$. This is indeed
the case if one puts together the results for $e^+e^-$, $pp$ and
$p\bar{p}$ and and $AA$ collisions for example, those of
\cite{qe+e-} for $e^+e^-$ collisions, together with those of
\cite{Info_new} for $p\bar{p}$ collisions and all results for
heavy-ion collisions, like \cite{ALQ,NUWW1,BD} and especially
\cite{C_V_Q} where such a trend was found when analyzing heavy-ion
events with different centrality (i.e., with different volumes
$V$).

\item As the parameter $q$ replaces in some sense the action of
many not yet identified dynamical factors, one expects that, with
such factors included, $q$ should diminish. This is precisely what
has been demonstrated in \cite{C_V_Q} analyzing transverse momenta
of pions produced in RHIC experiments by using a simple minded
Tsallis formula and an accordingly modified Hagedorn \cite{H}
approach which already contains in it some dynamics (based on a
special bootstrap hypothesis of resonances composed out of
resonances itd.). In the second case the values of $q-1$ found are
much smaller, but still remain nonzero indicating therefore the
existence of some residual additional  dynamic there.

\item As demonstrated in \cite{C_V_Q}, using a nonextensive
version of the statistical model allows us to describe data well
in the domain previously believed to be governed entirely be pure
jet physics. Deviations (i.e., dominance of truly hard collisions)
start at $p_T$ near $10$ GeV and further. It would mean that the
so called mini-jet region can probably also be investigated using
a nonextensive approach (what should be, however, checked in more
detail int the future).
\end{itemize}

\subsection{The whole phase space}
\label{sec:The whole...}

Already presenting results for the longitudinal phase space we
encountered multiplicity distributions of produced secondaries.
They involve the whole of phase space, both its longitudinal and
transverse components. However, as we have already stressed,
because $(q_L -1) >> (q_T - 1)$, the dominant role of the
longitudinal dynamic in establishing the actual number of produced
secondaries and its fluctuation from event to event is obvious. We
shall now discuss this problem in more detail.

Previous findings could be summarized in the following way:
knowing the amount of energy $W$ which is going to be transferred
to the produced secondaries (i.e., knowing the inelasticity $K$ of
reaction \cite{inel}) and the mean number of produced secondaries,
$\langle N\rangle$, and respecting the fact that they are
essentially distributed in the longitudinal phase space only, one
arrives, after using the information theory approach (cf.,
\cite{Info}), with the usual exponential distribution in $E =
\langle m_T\rangle \cosh y$. Additional information on the fact
that produced secondaries are distributed not according to a
Poisson distribution but rather according to NB distribution
characterized by parameter $k$ is enough to get $q$-exponential
distribution in $E$ with $q = 1 + 1/k$.

Now, it turns out that the opposite is also true, namely, as we
have shown in \cite{Q}, the fact that $N$ particles are
distributed in energy via $N$-particle Tsallis distribution
described by the nonextensivity parameter $q$ allows us to show
that their number distribution has to be of the NB type with $k =
1/(q-1)$. To illustrate this we first start with the derivation of
Poisson multiplicity distribution and then to compare it with the
corresponding derivation of the NB distribution \cite{Q}.

\subsubsection{Poisson multiplicity distribution}
\label{sec:P}

This distribution arises in a situation where in some process one
has $N$ independently produced secondaries with energies $\{
E_{1,\dots,N}\}$, each distributed according to the simple
Boltzmann  distribution:
\begin{equation}
f\left(E_i\right) = \frac{1}{\lambda}\cdot \exp\left( -
\frac{E_i}{\lambda}\right) \label{eq:f(E)}
\end{equation}
(where $\lambda =\langle E\rangle$). The corresponding joint
probability distribution is then given by:
\begin{equation}
f\left( \{ E_{1,\dots,N}\} \right) = \frac{1}{\lambda^N}\cdot
\exp\left( - \frac{1}{\lambda}\sum_{i=1}^N E_i\right)
.\label{eq:jointP}
\end{equation}
For independent energies $\{ E_{i=1,\dots,N}\}$ the sum $E =
\sum_{i=1}^N E_i$ is then distributed according to the following
gamma distribution,
\begin{equation}
g_N(E) = \frac{1}{\lambda (N-1)!} \cdot
\left(\frac{E}{\lambda}\right)^{N-1}\cdot \exp\left( -
\frac{E}{\lambda}\right), \label{eq:Egamma}
\end{equation}
distribuant of which is
\begin{equation}
G_N(E) = 1 - \sum_{i=1}^{N-1}\frac{1}{(i-1)!}\cdot
\left(\frac{E}{\lambda}\right)^{i-1}\cdot \exp( -
\frac{E}{\lambda}) . \label{eq:distribuant}
\end{equation}
Eq. (\ref{eq:Egamma}) follows immediately either by using
characteristic functions or by sequantially performing integration
of the joint distribution (\ref{eq:jointP}) and noticing that:
\begin{equation}
g_N(E) = g_{N-1}(E)\frac{E}{N-1} . \label{eq:E123ijkn}
\end{equation}
For energies such that
\begin{equation} \sum_{i=0}^N E_i \leq E \le
\sum_{i=0}^{N+1} E_i \label{eq:conditsumE}
\end{equation}
the corresponding multiplicity distribution has a Poissonian form
(notice that $E/\lambda = \langle N\rangle$):
\begin{eqnarray}
P(N) &=& G_{N+1}(E) - G_N(E) = \label{eq:Poisson} \\
&=& \frac{\left(\frac{E}{\lambda}\right)^N}{N!}\cdot \exp( -
\alpha E) = \frac{\langle N\rangle ^N}{N!}\cdot \exp( - \langle
N\rangle ).\nonumber
\end{eqnarray}
In other words, whenever we have variables $E_{1,\dots,N,N+1,\dots
}$ taken from the exponential distribution $f\left(E_i\right)$ and
whenever these variables satisfy the condition $\sum_{i=0}^N E_i
\leq E \le \sum_{i=0}^{N+1} E_i$, then the corresponding
multiplicity $N$ has a Poissonian distribution\footnote{Actually,
this is the method of generating Poisson distribution in the
numerical Monte-Carlo codes.}.

\subsubsection{Negative Binomial multiplicity distribution}
\label{sec:NB}

This distribution arises when in some process $N$ independent
particles with energies $\{ E_{1,\dots,N} \}$ which are now
distributed according to Tsallis distribution,
\begin{equation}
h\left( \{E_{1,\dots,N}\}\right) =
C_N\left[1-(1-q)\frac{\sum_{i=1}^N
E_i}{\lambda}\right]^{\frac{1}{1-q}+1 -N}, \label{eq:Enq}
\end{equation}
with normalization constant $ C_N$ given by
 \begin{eqnarray}
C_N &=& \frac{1}{\lambda^N}\prod_{i=1}^N[(i-2)q - (i-3)]
=\nonumber\\
& = & \frac{(q-1)^N}{\lambda^N}\cdot \frac{\Gamma\left( N +
\frac{2-q}{q-1}\right)}{\Gamma \left( \frac{2-q}{q-1}\right)}.
\label{eq:CNORMP}
\end{eqnarray}
It means that there are some intrinsic (so far unspecified but
summarily characterized by the parameter $q$) fluctuations present
in the system under consideration. In this case we do not know the
characteristic function for the Tsallis distribution, however,
because we are dealing here only with variables $\{
E_{i=1,\dots,N} \}$ occurring in the form of the sum, $E =
\sum_{i=1}^N E_i$, one can still sequentially perform integrations
of the joint probability distribution (\ref{eq:Enq}) and, noting
that (as before, cf. eq. (\ref{eq:E123ijkn}))
\begin{equation}
h_N(E) = h_{N-1}(E)\frac{E}{N-1} =
\frac{E^{N-1}}{(N-1)!}h\left(\{E_{1,\dots,N}\}\right),
\label{eq:qE123ijkn}
\end{equation}
we arrive at formula corresponding to eq. (\ref{eq:Egamma}),
namely
\begin{eqnarray}
h_N(E) &=& \frac{E^{(N-1)}}{(N-1)!\lambda^N}\times
\\
\times && \prod^N_{i=1}[(i-1)q - (i-3)]\left[ 1 -
(1-q)\frac{E}{\lambda}\right]^{\frac{1}{1-q}+1-N}\nonumber
\label{eq:Hq}
\end{eqnarray}
with distribuant given by
\begin{eqnarray}
H_N(E) &=& 1 - \sum_{j=1}^{N-1}  \tilde{H}_i(E)\qquad {\rm
where} \label{eq:qdistribuant}\\
\tilde{H}_i(E) &=& \frac{E^{i-1}}{(j-1)!\lambda^j}\times\nonumber
\\
\times && \prod_{i=1}^j \left[ (i-1)q - (i-3)\right]\left[ 1 -
(1-q)\frac{E}{\lambda} \right]^{\frac{1}{1-q}+1-j} .\nonumber
\end{eqnarray}
As before, for energies $E$ satisfying the condition given by eq.
(\ref{eq:conditsumE}), the corresponding multiplicity distribution
is equal to
\begin{equation}
P(N) = H_{N+1}(E) - H_N(E) \label{eq:qP(N)}
\end{equation}
and is given by the Negative Binomial distribution :
\begin{eqnarray}
P(N) &=& \frac{(q-1)^N}{N!}\cdot\frac{q-1}{2-q}\cdot
\frac{\Gamma\left(N+1+\frac{2-q}{q-1}\right)}{\Gamma \left(
\frac{2-q}{q-1}\right)}\times\nonumber\\
&\times & \left(\frac {E}{\lambda}\right)^N
\left[1-(1-q)\frac{E}{\lambda} \right]^{-N+\frac{1}{1-q}} =
\nonumber\\
&=& \frac{\Gamma(N+k)}{\Gamma(N+1)\Gamma(k)}\cdot \frac{\left(
\frac{\langle N\rangle}{k}\right)^N }{\left( 1 + \frac{\langle
N\rangle}{k}\right)^{N+k} } ,\label{eq:NBDfinal}
\end{eqnarray}
where the mean multiplicity and variance are, respectively,
\begin{eqnarray}
\langle N\rangle &=& \frac{E}{\lambda};\label{eq:meanNVarq}\\
Var(N) &=& \frac{E}{\lambda}\left[ 1 -
(1-q)\frac{E}{\lambda}\right] = \langle N\rangle + \langle
N\rangle^2 \cdot (q-1) .\nonumber
\end{eqnarray}
This distribution is defined by the parameter $k$ equal to:
\begin{equation}
k = \frac{1}{q-1} .\label{eq:k1}
\end{equation}
Notice that for $q \rightarrow 1$ one has $k\rightarrow \infty$
and $P(N)$ becomes a Poisson distribution, whereas for
$q\rightarrow 2$ one has $k\rightarrow 1$ and we are obtaining
geometrical distribution \footnote{Actually the parameter $k$ in
NB can be simply expressed by the correlation coefficient $\rho$
for the two-particle energy correlations, $ k = (\rho + 1)/\rho$,
see \cite{Q} for details.}.

\section{Further developments}
\label{sec:further...}

Let us now proceed a step further in Eq. (\ref{eq:HC}) by writing
it in the following form,
\begin{equation}
c_p\rho \frac{\partial T}{\partial t}\, =\, a\left( T' - T\right)
+ \eta f(u), \label{eq:visc}
\end{equation}
with a new term, $\eta f(u)$, which presents the effect of a
possible viscosity (with viscosity coefficient $\eta$) existing in
the system.  The function $f(u)$ contains terms dependent on the
velocity in the form of $\frac{\partial u_i}{\partial x_k} -
\frac{\partial u_k}{\partial x_l}$. Using as before $T'$ defined
by (\ref{eq:TTT}) we get an extension of Eq. (\ref{eq:T}):
\begin{eqnarray}
\frac{\partial T}{\partial t} + \left[ \frac{1}{\tau} +
\xi(t)\right] T &=& \frac{1}{\tau} T_0 + \eta f(u) \frac{1}{c_p
\rho} = \nonumber\\
&=& \frac{1}{\tau} \left[ T_0 + \frac{\eta \tau}{c_p \rho} f(u)
\right] . \label{eq:T_visc}
\end{eqnarray}
This equation leads to the Langevin equation resulting in
fluctuations of the temperature $T$ given in the same form of Eq.
(\ref{eq:FRES}) as before but with
\begin{equation}
\mu = \frac{1}{q-1}\left[ T_0 + \frac{\eta \tau}{c_p \rho} f(u)
\right] = \frac{T_{eff}}{q-1} .\label{eq:Teff}
\end{equation}
In this way previous $T_0  = \langle T\rangle$ has now been
replaced by a kind of {\it effective temperature}
\begin{equation}
T_{eff} = T_0 + \frac{\eta \tau}{c_p \rho} f(u) = T_0 +
\frac{\eta}{a} f(u) . \label{eq:Teff1}
\end{equation}
Introducing a kinetic coefficient of conductance $\nu = \eta/\rho$
and denoting $\kappa = c_p/c_V$, where $c_V$ is specific heat
under the constant volume for which $1/c_V = q - 1$, we have that
\begin{equation}
T_{eff}\, =\, T_0 + \frac{\nu \tau}{\kappa c_V} f(u)\, = \, T_0 +
(q-1)\frac{\nu \tau}{\kappa} f(u) \label{eq:Teff2}
\end{equation}
or, because $\tau D = q-1$, one can write this also as
\begin{equation}
T_{eff}\, =\, T_0 + (q-1)^2\frac{\nu }{\kappa D} f(u).
\label{eq:Teff3}
\end{equation}

\begin{figure}[h]
\begin{center}
\resizebox{0.45\textwidth}{!}{
  \includegraphics{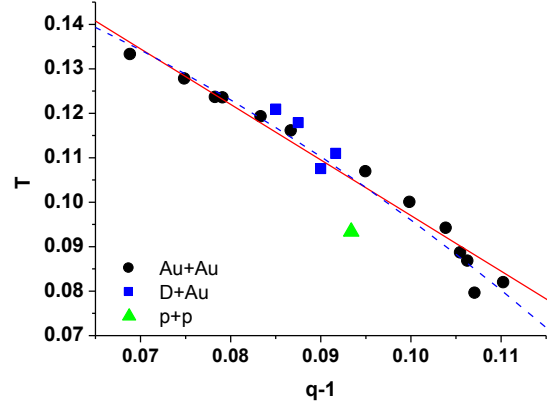}
} \vspace{-5cm} \caption{Dependence of temperature $T$ (in GeV) on
the parameter $q$ for production of negative pions in different
reactions. The solid line shows a linear fit to obtained results:
$T = 0.22 - 1.25 (q-1)$ (cf., Eq. (\ref{eq:Teff2})) and dashed
line shows the corresponding quadratic fit: $T = 0.17 - 7.5
(q-1)^2$ (cf.,  Eq. (\ref{eq:Teff3})).} \label{FigPi}
\end{center}
\end{figure}

\begin{figure}[h]
\vspace{-0.5cm}
\begin{center}
\resizebox{0.45\textwidth}{!}{
  \includegraphics{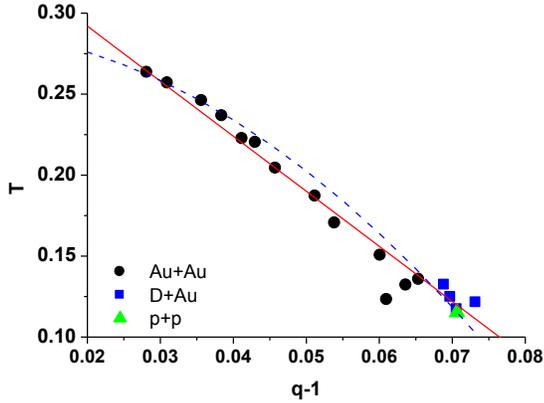}
} \vspace{-5cm} \caption{The same as in Fig. \ref{FigPi} but for
the produced antiprotons. The linear fit (solid line) is: $T =
0.36 - 3.4(q-1)$ whereas quadratic one (dashed line) is: $T = 0.29
- 35(q-1)^2$. } \label{FigAP}
\end{center}
\end{figure}

In \cite{BD} the transverse momentum spectra of pions and protons
and antiprotons produced in the interactions of P + P, D + Au and
Au + Au at $\sqrt{s_{NN}} = 200$ GeV at RHIC-BNL \cite{Data} were
analyzed using a nonextensive approach. Among other things they
found dependencies of the nonextensivity parameter $q$ and
temperature $T$ on the number of participants, $ N_{p}$ (i.e.,
number of nucleons taking part in a given $AA$ collision in the
production of secondaries). From them we have obtained a
dependence of $T$ on the parameter $q$ which are shown in Figs.
\ref{FigPi} and \ref{FigAP}. In all cases we find that $f(u) < 0$
and that $T$ seems to be linearly dependent on $q-1$.


\begin{figure}[h]
\begin{center}
 \resizebox{0.45\textwidth}{!}{
  \includegraphics{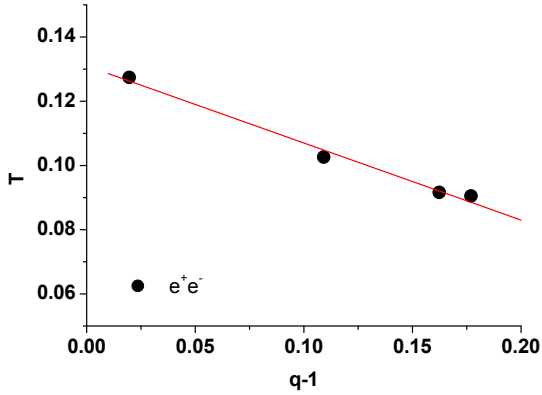}
} \vspace{-5.5cm} \caption{Dependence of temperature $T$ (in GeV)
on the parameter $q$ for the production of pions in $e^+e^-$
annihilation reactions. $T = 0.131 - 0.24(q-1)$ .} \label{Figee}
\end{center}
\vspace{-5mm}
\end{figure}

These results can be compared with old results for $e^+e^-$
annihilation reactions discussed some time ago in terms of
$q$-statistics in  \cite{qe+e-}. The $q$ dependence of the
temperature parameter $T$ which can be deduced from them is shown
in Fig. \ref{Figee}. Notice that now the temperature is lower and
depends only weakly on $q$.

Finally, let us discuss results on fluctuations of multiplicity
observed in heavy-ion collisions \cite{NA49}. They exhibit
non-monotonic changes as function of the number of participants
$N_p$ \cite{NA49}. Actually, also changes of $\langle N\rangle$
show nonlinear increase, though not so spectacular. Acting in the
spirit of our analysis here we can expect that
\begin{equation}
\frac{\frac{Var(N)}{<N>}-1}{<N>} = q-1,
\end{equation}
but now, with $T_{eff}$ we can show that
\begin{equation}
<N> = \frac{W}{T_{eff}}
\end{equation}
where $T_{eff}$ is given by Eq. (\ref{eq:Teff2}) and where $W$ is
the full accessible energy. We have therefore that
\begin{equation}
\frac{\langle N \rangle - n_0 N_p}{<N>} = c(q - 1).
\end{equation}
Here $n_0$ is the multiplicity in the single nucleon-nucleon
collision measured in the region of acceptance, $ c = - \frac{\nu
\tau}{\kappa}\frac{f(u)}{T_0}$ (notice that $c$ is positive
because, as was found from Figs. \ref{FigPi} and \ref{FigAP} ,
$f(u) < 0$).

\begin{figure}[h]
\vspace{-1.6cm}
 \resizebox{0.53\textwidth}{!}{\hspace{-2.5cm}
  \includegraphics{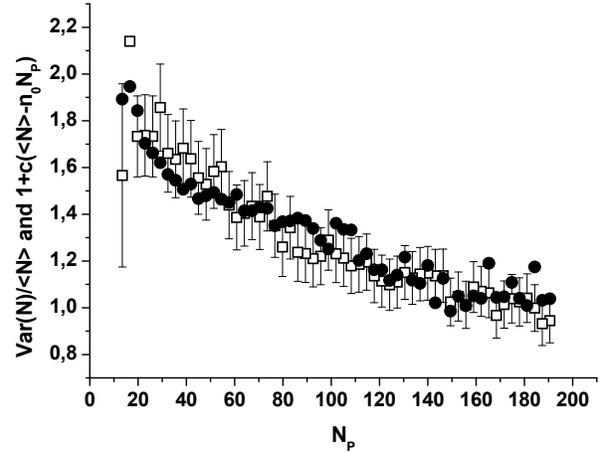}
} \vspace{-7.5cm} \caption{Comparison of $Var(N)/\langle N\rangle
$ versus $N_p$ (squares) with $1 + c (<N>-n_0 Np)$ vs $N_p$
(circles) (here $n_0 = 0.642$ and $c = 4.1$ . Data are for
negatively charged particles from $PbPb$ collisions as collected
by NA49 experiment \cite{NA49}. } \label{FigNA49-1}
\end{figure}

\begin{figure}[h]
\vspace{-1.2cm}
 \resizebox{0.53\textwidth}{!}{\hspace{-2.5cm}
  \includegraphics{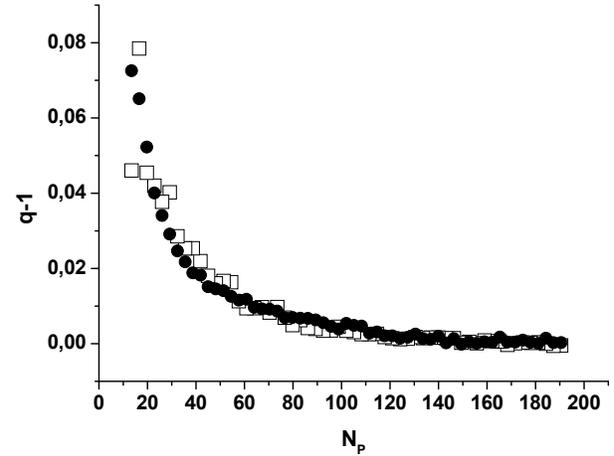}
} \vspace{-7.5cm} \caption{The same as in Fig. \ref{FigNA49-1} but
translated to $q-1$ vs $N_p$. Squares were obtained from
$Var(N)/\langle N\rangle$ vs $N_p$ and circles from $\langle
N\rangle$ vs $N_p$. As before $n_0 = 0.643$ and $c = 4.1$.}
\label{FigNA49-2}
\vspace{-0.3cm}
\end{figure}

In Figs. \ref{FigNA49-1} and \ref{FigNA49-2} we show what can be
extracted from $PbPb$ collision data taken by NA49 experiment
\cite{NA49}. As one can observe the data confirm our expectation
that dependencies of $Var(N)/\langle N\rangle$ and $\langle
N\rangle$ on the number of participants $N_p$ are, after
introducing the concept of $T_{eff}$, essentially the same. The
value of $n_0 = 0.642$ is also sensible, being only a little
greater than the multiplicity observed in $pp$ collisions when
calculated using the acceptance of the NA49 experiment. Notice
also that value of $c = 4.1$ obtained here for $PbPb$ collisions
is not far from the value $1.25/0.22 = 5.7$ obtained for data from
RHIC (i.e., for $AuAu$ collisions but at much higher energy) which
we have obtained in Fig. \ref{FigPi}.

We close this part by noticing that this problem is not trivial
since none of the known models for multiparticle production
processes describes $Var(N)/\langle N\rangle$ vs $N_p$ observed
experimentally \cite{NA49}. They are described only by some
specialized models addressing fluctuations, like the percolation
model \cite{FM1}, the model assuming inter-particle correlations
caused by the combination of strong and electromagnetic
interactions \cite{FM2} or the transparency, mixing and reflection
model \cite{FM3}. Actually, all those attempts were addressing
only $Var(N)/\langle N\rangle$ vs $N_p$ but not $\langle N\rangle$
vs $N_p$. From this perspective, results presented above in Figs.
\ref{FigNA49-1} and \ref{FigNA49-2} confirm the reasonableness of
the idea of $T_{eff}$ introduced in this Section. If one uses
$Var(N)/\langle N\rangle$ vs $N_p$ to obtain $q-1$ then it turns
out that the same value of $q-1$ describes also dependence of
$\langle N\rangle$ on $N_p$; this can only be using $T_{eff}$ and
this is because it depends on $q-1$.

\section{Remarks and summary}

Let us start with two remarks which are in order here:
\begin{itemize}
\item[$(i)$] Results which recall directly to Tsallis entropy were
obtained using the constraint $\sum p_i = 1$ and the formula
$\sum_i p_i^q A_i = \langle A\rangle _q $ for the $q$-expectation
values. On the other hand, there exists a formalism, which
expresses both the Tsallis entropy and the expectation values
using the so-called escort probability distributions
\cite{escort}: $P_i = p_i^q/\sum_i p_i^q$. However, as was shown
in \cite{Abe}, such an approach is different from the normal
nonextensive formalism because the Tsallis entropy expressed in
terms of the escort probability distributions has some difficulty
with the property of concavity. From our limited point of view, it
seems that there is no problem in what concerns practical,
phenomenological applications of nonextensivity as discussed in
the present work. Namely, using $P_i$ one gets distributions of
the type $c\left[1 - (1-q)x/l\right]^{q/(1-q)}$, which is, in
fact, {\it formally identical} with  $c\left[1 -
(1-Q)x/L\right]^{1/(1-Q)}$, provided we identify: $Q = 1 +
(q-1)/q$, $L=l/q$ and $c=(2-Q)/L=1/l$. The mean value is now
$\langle x\rangle = L/(3-2Q) = l/(2-q)$ and $0<Q<1.5$ (to be
compared with $0.5< q < 2$). Both distributions are identical and
the  problem, of which of them better describes data is
artificial.

\item[$(ii)$] One should be aware that there is still an ongoing
discussion on the meaning of the temperature in nonextensive
systems. However, the small values of the parameter $q$ deduced
from data in transverse phase space (where the connection with
thermodynamical approach makes sense, as discussed before) allow
us to argue that, to first approximation, $T$ can be regarded as
the hadronizing temperature in such a system. One must only
remember that in general what we study here is not so much the
state of equilibrium but rather some kind of stationary state. For
a thorough discussion of the temperature of nonextensive systems,
see \cite{T_q}.
\end{itemize}

With the above reservations in mind, we can summarize that, when
looking from the point of view of a statistical approach
\cite{Thermodynamics,QGP}, the power-law behavior of many
distributions observed in elementary and heavy ion collisions can
be traced back to the necessity of using the nonextensive version
of a statistical model (here taken in the form proposed by Tsallis
\cite{Tsallis}).

We interpret this as a sign of some intrinsic fluctuations present
in any hadronizing system, which were only recently to be
recognized as vital observable when searching for the production
of the QGP form of matter \cite{QGP}. In fact, a number of works
\cite{QGPex} have demonstrated the existence in such reactions of
event-by-event fluctuations of the average transverse momenta
$\langle p\rangle$ per event. The quantities considered were:
$Var\left(\langle p\rangle \right)/\langle \langle p\rangle
\rangle^2$ and $\langle \Delta p_i \Delta p_j\rangle/\langle
\langle p\rangle \rangle^2$. These quantities can be shown
\cite{Q} to be fully determined by $\omega$ as defined by Eq.
(\ref{eq:omega1}), i.e., by fluctuations of the temperature $T$ of
hadronizing system - a vital observable when searching for
QGP\footnote{Generally speaking, an analysis of transverse momenta
$p_T$ alone indicates very small fluctuations of $T$. On the other
hand, as reported in \cite{MR}, the measured fluctuations of
multiplicities of produced secondaries are large (i.e.,
multiplicity distributions are substantially broader than
Poissonian).}. In fact, when considering  the case of $N_{ev}$
events with $N_k$ particles in the $k^{\rm th}$ event, one has
that
\begin{equation}
\frac{Var(\langle p\rangle)}{\langle \langle p \rangle\rangle^2} =
\frac{Var(T)}{\langle T\rangle^2} = \omega. \label{eq:C555}
\end{equation}
where
\begin{eqnarray}
\langle \langle p \rangle \rangle &=&\frac{1}{N_{ev}}\,
\sum_k^{N_{ev}}\langle p\rangle_k;\quad {\rm with}\quad \langle
p\rangle_k = \frac{1}{N_k}\sum_i^{N_k}p_i , \label{eq:llprr}
\end{eqnarray}
This is what we have shown in the last part od Section
\ref{sec:further...}. This is the problem which needs further
investigations.

We close with some remarks:
\begin{itemize}
\item Although our original investigations presented here were
based on the notion of Tsallis entropy (usually with the help of
information theory) one must mention that one can also get Tsallis
distribution without resorting to a Tsallis entropy altogether
(see, for example, \cite{Select}).

\item The other way to get a Tsallis distribution from some
general thermodynamical considerations was presented in \cite{GE}.
It is based on allowing a linear dependence of the temperature $T$
on energy, $T = T_0 + (q-1) E$. Here temperature is not
fluctuating. Actually, if one would like to follow this approach
and to have Tsallis distribution with $T_{eff}$ discussed in
Section \ref{sec:further...} one should write $T = T_0 +
(q-1)\cdot {\rm const} + (q-1) E$. Then $T = T_0$ only would
result in $\exp \left( - E/T_0 \right)$, $T = T_0 (q-1) E$ would
result in the usual Tsallis distribution $\exp_q\ \left( -
E/T_0\right)$, $T = T_{eff} = T_0 +(q-1)\cdot {\rm const}$ would
give $\exp\left( -E/T_{eff}\right)$ and, finally, $T = T_{eff} =
T_0 + (q-1)\cdot {\rm const} + (q-1) E$ would give $\exp_q \left(
-E/T_{eff}\right)$.

\item  Notice that for $x >> \lambda/(q-1)$ Tsallis distribution
becomes a pure power low and loses its dependence on the scale
$\lambda$: $ f(x)\sim [1-(1-q)x/\lambda]^{1/(1-q)} \longrightarrow
x^{1/(1-q)}$.

\item Instead of using an intrinsic fluctuations one can also
obtain a power law distribution by using the notion of
self-organized criticality \cite{SOC} in a cascade processes (cf.,
\cite{q_approaches,RWW}).

\item Another interesting possibility, not yet fully explored, is
that, as shown in \cite{WWqnet}, one can formulate a description
of the so called stochastic networks using nonextensive
information theory based on Tsallis statistics. Using this
approach one can then demonstrate \cite{WWhnet} that hadron
production viewed as formation of a specific stochastic network
can explain in a natural way the power-law distributions of
transverse mass spectra of pions found in \cite{GG}.

\item In the string models of production of hadrons the natural
distribution in $p_T$ is $\exp \left( -\pi m^2_T/\kappa\right)$
rather than $\exp\left( m_T/T\right)$ (where $\kappa$ is string
tension) really observed. However, if one allows for the gaussian
fluctuations of the parameter $\kappa$ (characterized by parameter
$\langle \kappa ^2\rangle$ which can be connected with the
fluctuations in the QCD vacuum) then the first form is transformed
into the second one with $T = \sqrt{\langle
\kappa^2\rangle/(2\pi)}$ (i.e., in this approach parameter $T$
characterizes rather the properties of the QCD vacuum than those
of hadrons) - see \cite{Bialas}.

\item Finally, recall that when applied to the hydrodynamical
model of multiparticle production the nonextensivity approach
converts the usual nonviscous hadronic fluid into the viscous one
preserving, however, the usual linear flow equations (albeit now
given in a nonextensive form) \cite{Osada2007}.

\end{itemize}

\begin{acknowledgement}
Partial support (GW) of the Ministry of Science and Higher
Education under contracts 1P03B02230 is acknowledged.
\end{acknowledgement}


\begin{thebibliography}{}

\bibitem{Thermodynamics} R. Hagedorn, Nuovo Cim. (Suppl.) {\bf 3}, 147
                         (1965), Nuovo Cim. A {\bf 52}, 64 (1967) and
                         Riv. Nuovo Cim. {\bf 6} (1983);
                         C. Geich-Gimbel, Int. J. Mod. Phys. A {\bf 4}, 1527 (1989).
                         U. Heinz, J. Phys. G {\bf 25}, 263 (1999) 263;
                         F. Becattini, Nucl. Phys. A {\bf 702}, 336 (2002),
                         F. Becattini, G. Passaleva, Eur. Phys. J. C {\bf 23},
                         551 (2002); W. Broniowski, A. Baran, W. Florkowski,
                         Acta Phys. Polon. B {\bf 33}, 4325 (2002) 4235.

\bibitem{QGP} Cf., QM2008 proceedings, J. Phys. G {\bf 35} (10)  (2008);
              see also: B.M\"uller, Nucl. Phys. A {\bf774}, 433 (2006);
              M. Gyulassy and L. McLerran, ibid. A {\bf 750}, 30 (2005); I.
              Vitev, Int. J. Mod. Phys. A {\bf 20}, 3777 (2005); R. D. Pisarski, Braz.
              J. Phys. {\bf 36}, 122 (2006) and references therein.

\bibitem{LVH} L. Van Hove, Z. Phys. C {\bf 21}, 93 (1985)
              and C {\bf 27}, 135 (1985).

\bibitem{dynamics} B. M\"uller and J. L. Nagle, Annu. Rev. Nucl. Part. Sci. {\bf 56}, 93
                   (2006) and references therein.

\bibitem{Info} G. Wilk, Z. W\l odarczyk, Phys. Rev. D {\bf 43}, 794
               (1991); O.V.Utyuzh, G.Wilk and Z.W\l odarczyk, Acta Phys. Hung. A -
               Heavy Ion Phys. {\bf 25}, 65 (2006).

\bibitem{Info_new} F.S. Navarra, O.V. Utyuzh, G. Wilk and Z. W\l odarczyk,
                   Physica A {\bf 340}, 467 (2004).

\bibitem{Measures} C. Arndt, \textit{Information measures - Information and its Description
                   in Science and Engineering}, Springer, 2004;
                   T.I.J. Taneja, \textit{Generalized Information Measures and Their
                   Applications}, on-line book,
                    http://www.mtm.ufsc.br/taneja/ book/book.html.

\bibitem{Tsallis} C. Tsallis, J. Stat. Phys. {\bf 52}, 479 (1988),
                  Braz. J. Phys. {\bf 29}, 1 (1999),
                  Physica A {\bf 340}, 1 (2004) and
                  Physica A {\bf 344}, 718 (2004) and references therein.
                  See also {\it Nonextensive Statistical Mechanics and its Applications},
                  S. Abe and Y. Okamoto (Eds.), Lecture Notes in Physics
                  {\bf LPN560}, Springer (2000) and {\it
                  Nonextensive Entropy - interdisciplinary
                  applications}, M. Gell-Mann and C. Tsallis
                  (Eds.), (a volume in the Santa Fe Institute Studies in the Science of
                  Complexity), Oxford University Press (2004).
                  For an updated bibliography on this subject see
                    http://tsallis.cat.cbpf.br/biblio.htm;
                  C. Tsallis, M. Gell-Mann, Y. Sato, {\it Extensivity and entropy production},
                  Europhysics News {\bf 36}, 186 (2005) 186 (in: J.P. Boon, C. Tsallis
                 (Eds.), {\it Nonextensive Satistical Mechanics: New Trends, New
                  Perspectives}, Europhysics News (special issue), 2005.

\bibitem{Correlations} T. Kodama, H.-T. Elze, C.E. Augiar and T. Koide, Europhys. Lett.
                       {\bf 70}, 439 (2005); T. Kodama, J. Phys. G
                       {\bf 31}, S1051 (2005).

\bibitem{Fractal} T. La\u{s}tovi\u{c}ka, Eur. Phys. J. C {\bf 24}, 529
                  (2002).

\bibitem{q_interpretation} G. Wilk and Z. W\l odarczyk, Phys. Rev. Lett. {\bf 84}, 2770
                           (2000).

\bibitem{q_interpretation_1} G. Wilk and Z. W\l odarczyk, Chaos, Solitons and Fractals
                             {\bf 13/3}, 581 (2001).

\bibitem{NUWWx} F.S. Navarra, O.V. Utyuzh, G. Wilk, and Z. W\l odarczyk,
                Nuovo Cimento Soc. Ital. Fis., C {\bf 24}, 725 (2001).

\bibitem{SuperS} C. Beck, E.G.D. Cohen, Physica A {\bf 322}, 267 (2003);
                 F. Sattin, Eur. Phys. J. B {\bf 49}, 219 (2006).

\bibitem{UWWpT} O.V. Utyuzh, G. Wilk and Z. W\l odarczyk, J. Phys. G {\bf 26}, L39
                (2000).

\bibitem{q_approaches}  G. Wilk and Z. W\l odarczyk, Physica A {\bf 305}, 227 (2002);

\bibitem{q_approaches1}  M. Rybczy\'nski, Z. W\l odarczyk and G. Wilk,
                         Nucl. Phys. (Proc. Suppl.) B {\bf 122}, 325 (2003);

\bibitem{q_approaches2} T. Osada, O.V. Utyuzh, G. Wilk and Z. W\l odarczyk, Europ. Phys.
                         J. B {\bf 50}, 7 (2006).

\bibitem{NUWW} F.S. Navarra, O.V. Utyuzh, G. Wilk and Z. W\l odarczyk, Phys. Rev. D {\bf 67},
               114002 (2003).

\bibitem{NUWW1} F.S. Navarra, O.V. Utyuzh, G. Wilk and Z. W\l odarczyk, Physica A
                {\bf 344}, 568  (2004);

\bibitem{C_V_Q} M. Biyajima, M. Kaneyama, T. Mizoguchi and G. Wilk,
                Eur. Phys. J. C {\bf 40}, 243 (2005); M. Biyajima, T. Mizoguchi, N. Nakajima,
                N. Suzuki, and G. Wilk, Eur. Phys. J. C {\bf 48}, 593 (2006).

\bibitem{Q} G. Wilk and Z. W\l odarczyk, Physica A {\bf 376}, 279 (2007).

\bibitem{qe+e-} I. Bediaga, E.M. Curado and J.M.de Miranda, Physica A {\bf 286}, 156
                (2000).

\bibitem{ALQ} W.M. Alberico, A. Lavagno and P. Quarati, Eur. Phys. J. C {\bf 12}, 499 (2000);

\bibitem{Wibig} T. Wibig and I. Kurp,  J. High Energy Phys. {\bf 12}, 039
                (2003).

\bibitem{q_approaches_others} A. Lavagno, Physica A {\bf 305}, 238 (2002);
                              W. M. Alberico, P. Czerski, A. Lavagno, M. Nardi, and V. Som\'a,
                              Physica A {\bf 387}, 467 (2008).

\bibitem{AK} C.E. Aguiar and T. Kodama, Physica A {\bf 320}, 371 (2003).

\bibitem{BD} B. De, S. Bhattacharyya, G. Sau and S.K. Biswas, Int.
             J. Mod. Phys. E \textbf{16} (2007) 1687.

\bibitem{SR} T. Sherman and J. Rafelski,  Lecture Notes in Physics {\bf 633}, 377 (2004).

\bibitem{Biro} T.S. Bir\'{o} and G. Purcsel, Phys. Rev. Lett. {\bf 95}  162302
               (2005); Phys. Lett. A {\bf 372}, 1174 (2008). See
               also: T.S. Bir\'{o}, {\it Abstract composition rule for relativistic
               kinetic energy in the thermodynamical limit},
               arXiv: 0809.4675 [nucl-th].

\bibitem{BiroK} T.S. Bir\'{o} and G. Kaniadakis, Eur. Phys. J. B
                 {\bf 50}, 3 (2006) and references therein.

\bibitem{Beck} C. Beck, Physica A {\bf 286}, 164 (2000).

\bibitem{q_Cosmic} G. Wilk and Z. W\l odarczyk, Nucl. Phys. B
                        (Proc. Suppl.) {\bf 75A},  191 (1999).

\bibitem{Cosmic} G. Wilk and Z. W\l odarczyk, Phys. Rev. D {\bf 50}, 2318 (1994).

\bibitem{sigmafluct} H. Heiselberg {\it et al.}, Phys. Rev. Lett. {\bf 67}, 2946
                     (1991); B. Bl\"attel {\it et al.}, Phys. Rev. D {\bf 47}, 2761
                     (1993); L. Frankfurt, V. Guzey and M. Strikman, J. Phys. G
                     {\bf 27}, R23 (2001).

\bibitem{newsigmafluct} L. Frankfurt, M. Strikman, D. Treleani and C. Weiss,
                        {\it Evidence for color fluctuations in
                        the nucleon in high-energy collisions},
                        arXiv:0808.0182[hep-ph] (and in
                        preparation).

\bibitem{LLH} L.D. Landau and I.M. Lifschitz, \textit{Course of Theoretical
               Physics: Hydrodynamics}, Pergamon Press, New York 1958
               or {\it Course of Theoretical Physics: Mechanics of
               Continous Media}, Pergamon Press, Oxford 1981.

\bibitem{BiroJ} T.S. Bir\'{o} and A. Jakov\'ac, Phys. Rev. Lett. {\bf 94}, 132302 (2005).

\bibitem{inel} Y.-A. Chao, Nucl. Phys. B {\bf 40}, 475 (1972);
               Y.M. Shabelski, R.M. Weiner, G. Wilk, and Z. W\l odarczyk,
               J. Phys. G {\bf 18}, 1281 (1992);
               F.O. Dur\~aes, F.S. Navarra and G. Wilk, Braz. J. Phys. {\bf 35}, 3 (2005).

\bibitem{Datapp1} C. De Marzo et al., Phys. Rev. D {\bf 26}, 1019
                   (1982) and  D {\bf 29}, 2476 (1984);
                   R. Baltrusaitis et al.,Phys. Rev. Lett. {\bf
                   52},1380 (1993); F. Abe et al., Phys. Rev. D
                   {\bf 41}, 2330 (1990).

\bibitem{Data5} R. Barate, et al., (ALEPH Collab.), Phys. Rep. {\bf 294}, 1
                (1998).

\bibitem{DataAA} S.V. Afanasjev et al. (NA49 Collab.), Phys. Rev. C {\bf 66}, 054902 (2002).

\bibitem{Data4} B.B. Beck et al. (PHOBOS Coll.), Phys. Rev. Lett. {\bf 91}, 052303
                (2003).

\bibitem{partitionT} T.T. Chou and C.N. Yang, Phys. Rev. Lett. {\bf 54}, 510 (1985);
                     Phys. Rev. D {\bf 32}, 1692 (1985).

\bibitem{Shih} P. Carruthers and C.S. Shih, Int. J. Mod. Phys. A {\bf 2}, 1447 (1986).

\bibitem{Datapt} C. Albajar et al. (UA1 Collab.) Nucl. Phys. B
                  {\bf 335}, 261 (1990).

\bibitem{DatapT} T. Alber {\it et al} (NA35 Collaboration), Eur. Phys. J. C {\bf 2}, 643
                 (1998).

\bibitem{Capacity} L.D. Landau and I.M. Lifschitz, \textit{Course of Theoretical Physics:
                   Statistical Physics}, Pergmon Press, New York 1958.

\bibitem{GEN} L. Stodolsky, Phys. Rev. Lett. {\bf 75}, 1044 (1995);
               S. Mr\'{o}wczy\'{n}ski, Phys. Lett. B {\bf 430}, 9 (1998);
               E.V. Shuryak, Phys. Lett. B {\bf 423}, 9 (1998).

\bibitem{QCD} M. Stephanov, K. Rajagopal and E. Shuryak, Phys. Rev.
              Lett. {\bf 81}, 4816 (1998) and Phys. Rev. D {\bf 60}, 114028
              (1999); S. Mr\'{o}wczy\'{n}ski, Phys. Rev. C  {\bf 57}, 1518 (1998).

\bibitem{FLUCT} T.C.P. Chui, D.R. Swanson, M.J. Adriaans, J.A. Nissen
                and J.A. Lipa, Phys. Rev. Lett. {\bf 69}, 3005 (1992);
                C. Kittel, Physics Today {\bf 5}, 93 (1988);
                B.B. Mandelbrot, Physics Today {\bf 42}, 71 (1989);
                H.B. Prosper, Am. J. Phys. {\bf 61}, 54 (1993);
                G.D.J. Phillies, Am. J. Phys. {\bf 52}, 629 (1984).

\bibitem{KM} K. Kadaja and M. Martinis, Z. Phys. C {\bf 56}, 437
             (1992).

\bibitem{H} R. Hagedorn, Nuovo Cim. Suppl. {\bf 3}, 147 (1965);
            Nuovo Cim. A {\bf 52}, 64 (1967) and CERN Report 71-12,
            1971.

\bibitem{Data} S.S. Adler {\it et al.} (PHENIX Coll.), Phys. Rev. C
               \textbf{69} (2004) 034909;
               J. Adams {\it et al.} (STAR Coll.) Phys. Lett. B
               \textbf{616}, 8 (2005) and Phys.Lett. B \textbf{637}, 161 (2006.

\bibitem{NA49} C. Alt {\it et al.} (NA49 Collaboration), Phys. Rev. C {\bf 75}, 064904
               (2007).

\bibitem{FM1} E. G. Ferreiro, F. del Moral, and C. Pajares, Phys. Rev. C {\bf 69},
              034901 (2004).

\bibitem{FM2} M. Rybczy\'nski and Z. W\l odarczyk, J. Phys. Conf. Ser. {\bf 5},
              238 (2005).

\bibitem{FM3} M. Ga\'zdzicki and M. Gorenstein, Phys. Lett. B {\bf 640}, 155
              (2006) (2006).

\bibitem{escort} C. Tsallis, R.S. Mendes and A.R. Plastino, {\sl Physica}
                 {\bf A261}, 534 (1998).

\bibitem{Abe} S. Abe, Phys. Lett. A {\bf 275}, 250 (2000).

\bibitem{T_q}  S. Abe, Physica A {\bf 368}, 430 (2006).

\bibitem{QGPex} W. Broniowski, B. Hiller, W. Florkowski, and P. Bo\.zek, Phys.
                Lett. B {\bf 635}, 290 (2006); F. Jinghua et al, Phys. Rev. C {\bf 72},
                017901 (2005); J. Adams et al. (STAR Collab.), Phys. Rev. C {\bf
                72}, 044902 (2005); K. Adcox et al., (PHENIX Collab.), Phys. Rev. C {\bf
                66}, 024901 (2002).

 \bibitem{MR} M. Rybczy\'nski et al. (NA49 Collab.), J. Phys. Conf.
              Ser. {\bf 5}, 74 (2005).


\bibitem{Select} G.Wilk and Z.W\l odarczyk, AIP Conference Proceedings {\bf 965},
                 76 (2007)

\bibitem{GE} M.P. Almeida, Physica A {\bf 300}, 424 (2001) and {\bf 325}, 426 (2003).

\bibitem{SOC} Fu Jinghua, Meng Ta-chung, R. Rittel and K. Tabelow,
             Phys. Rev. Lett. {\bf 86}, 1961 (2001).

\bibitem{RWW} M. Rybczy\'nski, Z. W\l odarczyk and G. Wilk,
              Nucl. Phys. (Proc. Suppl.) B {\bf 97}, 81 (2001).

\bibitem{WWqnet} G. Wilk and Z. W\l odarczyk, Acta Phys. Polon. B {\bf 35},
                 871 (2004).

\bibitem{WWhnet} G. Wilk and Z. W\l odarczyk, Acta Phys. Polon. B {\bf 35},
                 2141 (2004).

\bibitem{GG} M. Ga\'zdzicki and M.I. Gorenstein, Phys. Lett. B {\bf 517}, 250 (2001).

\bibitem{Bialas} A. Bia\l as, Phys. Lett. B {\bf 466}, 301 (1999).

\bibitem{Osada2007} T.~Osada and G.~Wilk, Phys. Rev. C {\bf 77}, 044903
                    and Prog. Theor. Phys. Suppl. {\bf 174}, 168
                    (2008); see also {\it Dissipative or just Nonextensive hydrodynamics? -
                    Nonextensive/Dissipative correspondence -},
                    arXiv:0805.2253[nucl-phys] (to be published in Indian Journal of
                    Physics).


\end{thebibliography}
\end{document}